\documentclass[usegraphicx,usedcolumn,usenatbib,epsf,graphics]{mn2e}
\usepackage{rotating}
\newif\ifAMStwofonts
\AMStwofontstrue

\ifCUPmtlplainloaded \else
  \ifAMStwofonts \else 
    \def\upi{\pi}
    \def\umu{\mu}
    \def\upartial{\partial}
  \fi
\fi

\title[The infrared Hourglass cluster in M8]
{The infrared Hourglass cluster in M8\thanks{Based in part on observations
with NASA/ESA {\em Hubble Space Telescope} obtained from the archive at the
Space Telescope Science Institute, which is operated by the Association of
Universities for Research in Astronomy, Inc., under NASA contract NAS
5-26555.}
\thanks{Based in part on observations obtained at European Southern
Observatory, La Silla, Chile.}}

\author[Arias et al.]
{J.I. Arias$^1$\thanks{Fellow of CONICET, Argentina.},
R.H. Barb\'a$^2$\thanks{Member of Carrera del Investigador Cient\'{\i}fico,
CONICET, Argentina.},
J. Ma\'{\i}z Apell\'aniz$^3$\thanks{Affiliated with the Space Telescope 
Division of the European Space Agency, ESTEC, Noordwijk, Netherlands.}, 
N.I. Morrell$^{4}${\Large{\S}}, M. Rubio$^5$\\
$^1$ Facultad de Ciencias Astron\'omicas y Geof\'{\i}sicas, Universidad Nacional de La Plata, Paseo del Bosque S/N, B1900FWA La Plata, Argentina\\
$^2$ Departmento de F\'{\i}sica, Universidad de La Serena, Benavente 980, 
La Serena, Chile\\
$^3$ Space Telescope Science Institute, 3700 San Martin Drive, Baltimore, MD 21218\\
$^4$ Las Campanas Observatory, Carnegie Observatories, Casilla 601, 
La Serena, Chile\\
$^5$ Departamento de Astronom\'{\i}a, Universidad de Chile, Casilla 36-D,
Santiago, Chile\\
}

\pubyear{2003}

\begin{document} 

\maketitle

\begin{abstract}

A detailed study of the Hourglass Nebula in the M8 star forming 
region is presented. The study is mainly based on recent subarcsec-resolution 
$JHK_s$ images taken at Las Campanas Observatory and complemented with
archival {\em HST} images and longslit spectroscopy retrieved from the ESO
Archive Facility.
Using the new numerical code CHORIZOS, we estimate the distance to 
the earliest stars in the region to be 1.25~kpc.
Infrared photometry of all the sources detected in the field is given.
From analysis of the $JHK_s$ colour-colour diagrams, we find that an important
fraction of these sources exhibit significant infrared excess.
These objects are candidates to be low- and intermediate-mass 
pre-main sequence stars. 
Based on {\em HST} observations, the spatial distribution of gas, dust and 
stars in the region is analyzed.
The morphological analysis of these images also reveals 
a rich variety of structures related to star formation (proplyds,
jets, bow shocks), similar to those observed in M16 and M42,
along with the detection of the first four Herbig-Haro objects in the region.
Furthermore, a longslit spectrum obtained with NTT confirms the 
identification of one of them (HH\,870) in the core of the Hourglass nebula, 
providing the first direct evidence of active star formation by accretion
in M8.
\end{abstract}


\section{Introduction}
The Lagoon Nebula (Messier 8 = NGC~6523-NGC~6530) is an extended 
H\,{\sc ii} region mainly ionized by two O-type stars, 
9\,Sagitarii [O4\,V((f))] and HD~165052 (O6.5\,V + O7.5\,V).
It is embedded within a molecular cloud which extends to the star 
cluster NGC~6530. 
This very young open cluster is believed to be the starting point
for a sequential star formation process (Lightfoot et al. 1984)
which is still active in the region.
Within M8's core lies a  
distinctive bipolar nebula called the Hourglass, a blister-type H\,{\sc ii} 
region which has been produced by the O7.5\,V star Herschel~36 (Her\,36).
The Hourglass, which is about 15'' EW $\times$ 30'' NS in size, is
believed to be an ionized cavity in an inhomogeneous clumpy molecular cloud.
The Hourglass region also harbours the ultracompact H\,{\sc ii} region 
G5.97-1.17 as well as a number of infrared sources, first observed by Allen 
(1986), which may form a cluster of very young hot stars, analogous to the 
Orion Trapezium (Allen 1986).
Narrow band optical imaging with {\em HST} reveals that G5.97-1.17 
could be a young star surrounded by a circumstellar disk that is being 
photoevaporated by Her\,36, similar to the so-called proplyds
seen in the Orion Nebula (Stecklum et al. 1998). 
Furthermore, strong 
H$_2$ line emission is produced from around the Hourglass, 
showing a morphology very similar to that of the CO J=3-2 distribution 
in the region (White et al. 1997; Burton 2002). 
This fact and the detection of a jet-like feature extending from Her\,36 
in {\em HST} images (Stecklum et al. 1995) suggest there might be 
a molecular outflow in the core of M8 (Burton 2002).
Recently, Rauw et al. (2002) reported X-ray emission from Her\,36 
as well as probably diffuse X-ray emission from the Hourglass region 
that might reveal a bubble of hot gas produced by the interaction of the 
stellar wind of Her\,36 with the denser part of the molecular cloud.

In this paper, new high-resolution near-infrared images and photometry of the 
field surrounding the Hourglass are presented and compared with archival
{\em HST} emission-line images.  
All these data provide strong evidence that 
the core of M8 is an important region of active 
star formation and pre-main sequence stellar evolution.

\section{Observations and Data Reduction}

\subsection{Infrared Images}

$J$, $H$ and $Kshort$ ($K_s$) images of the Hourglass region
were obtained on September 26 1999, using the near-infrared camera IRCAM,
attached to the 2.5-m Du Pont Telescope at Las Campanas Observatory
(LCO), Chile. IRCAM was equipped with a NICMOS III $256 \times 256$
array (Persson et al. 1992) and has a pixel scale of 0\farcs35 $px^{-1}$.
The seeing during the observations was typically of 0\farcs8-0\farcs9
giving an optimum sampling.
In each band, five partially overlapping frames were taken, covering
a total area of 135''$\times$139'' 
centered approximately at the position of Her\,36.
The total on-source times were 360 s in $K_s$, 500 s in $H$ and 600 s in $J$.
The frames were combined (median-averaged) after being linearised,
dark- and sky-substracted, flat-fielded and cleaned of bad pixels, 
using SQIID processing routines layered in 
IRAF\footnote{IRAF is distributed by NOAO, operated by AURA, Inc., 
under agreement with NSF.}.
Several standard stars selected from Persson et al. (1998) were
observed during the night; these fields were reduced following a similar
procedure.
The combined (final) images of the Hourglass were then 
shifted to align stars in each frame.

\subsection{{\em HST}/WFPC2 Imaging}

A partial {\em HST}/WFPC2 coverage of the Hourglass Nebula is available from
the public database. These
images include the brightest part of the nebula and the star Her\,36, 
both centred in the Planetary Camera chip.  
The orientation of the images (P.A.= 56\fdg0919) allow to cover the west 
part of the nebula and they correspond to about $2/3$ of the field observed in 
our infrared images. The WFPC2 data set 
(Proposal ID. 6227) was obtained in 1995 July 13, 14, and September 16,
using several narrow band and broad band filters. 
These observations were retrieved from the {\em HST} Archive and  
processed using the ``on-the-fly'' pipeline.
In this work, we use images  obtained in F487N (H$\beta$), 
F656N (H$\alpha$), F658N ([N\,{\sc ii}]), 
F673N ([S\,{\sc ii}]) narrow band filters, and  F547M ($v$),  
F814W (Johnson $I$) broad band filters.
The filter characteristics are described in the WFPC2 Instrument Handbook.
The first steps of data reduction were straightforward, using IRAF and STSDAS
tasks to do cosmic ray cleaning, and integrating the four individual WFPC2 CCD
images into a single mosaic. 
In order to obtain flux calibrated "pure nebular" H$\beta$ and H$\alpha$ 
images, we followed the procedure given by O'Dell \& Doi (1999).
A "pure" H$\beta$ calibrated image was derived from the F487N image,
using the F547M images as continuum. A "pure" H$\alpha$ calibrated image was
obtained from F656N "line" image, considering the contamination of
[N\,{\sc ii}] emission lines through the F658N filter, and again using the
F547M as continuum image. 
In the case of F673N filter used to derive "pure"
[S\,{\sc ii}] calibrated emission line image, the continuum was interpolated
between the F547M and F814W images and  the zero-points determined by
{\tt synphot} (STScI 1999).

\subsection{Long-slit Echelle Spectroscopy}

Two long-slit spectra obtained with ESO Multi-Mode Instrument (EMMI) on
NTT were retrieved from the ESO Archive Facility. The spectra
were obtained on July 31st 2000 in the REMD
(red medium dispersion) configuration  with echelle (\#10) grating.
The order containing H$\alpha$ and [N\,{\sc ii}] emission lines was isolated
using the filter Ha\#596. The detector was a $2048 \times 2048$ Tek CCD
with 24-$\mu$m pixels, given a pixel scale of 
0\farcs265 $\times$ 0.041\,\AA\,
and a spatial extent along the slit of $330''$. The spectra were obtained
with a $P.A.=0^{\circ}$. The total exposure time was 1800\,s.

\section{Results}

\begin{figure*}
\includegraphics[width=180mm]{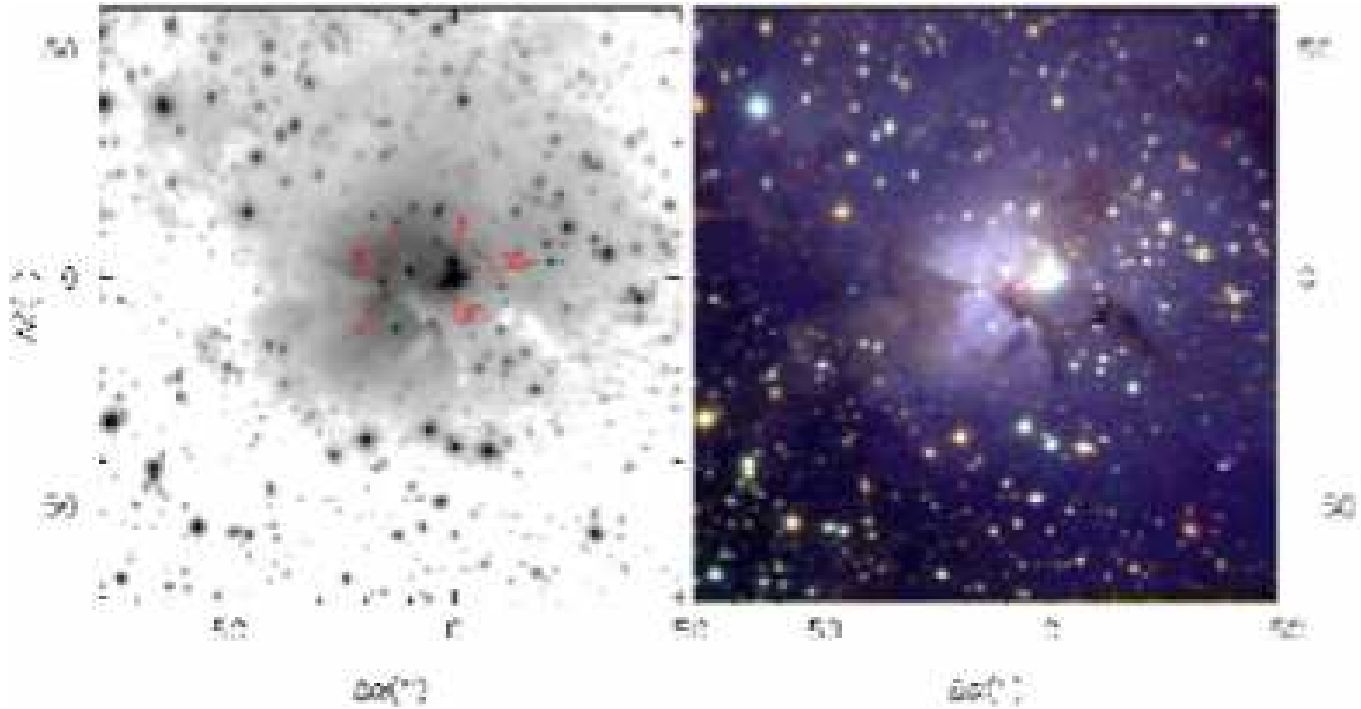}

\caption{Left-hand panel: $K_s$ image of the field surrounding the Hourglass 
Nebula. Some features of interest are labelled: the ionizing O7.5\,V star 
Herschel\,36 (H36), the infrared sources KS\,1-KS\,4 and the ultracompact 
H\,{\sc ii} region G5.97-1.17 (UC). 
Right-hand panel: three-colour near-infrared image of the same field. The 
large majority of sources in this image are not visible at optical 
wavelengths.}
\label{imagen}
\end{figure*}

Figure~\ref{imagen} (left-hand panel) shows the $K_s$
frame of the field surrounding the Hourglass Nebula.
Figure~\ref{imagen} (right-hand panel) is a $JHK_s$ colour composite image, 
produced combining the $J$ (1.25 $\mu$), $H$ (1.65 $\mu$)
and $K_s$ (2.2 $\mu$) images as 
{\em blue}, {\em green}, and {\em red} channels, respectively. 
This {\em false-colour} picture shows the effects of extinction, the
differences in colour of the stars mainly being due to differences in
reddening. This figures confirm the existence of an infrared star cluster 
around Her\,36, as was recently reported by Bica et al. (2003).

\subsection{Distance}

Most previous investigators estimated the distance to NGC~6530 as 1.8~Kpc
($V_0 - M_V = 11.25-11.3$)(van Altena \& Jones 1972; Sagar \& Joshi 1978; 
van den Ancker et al. 1997). A smaller value was however obtained by 
Kilambi (1977) ($V_0 - M_V = 10.7$).
More recently, Sung et al. (2000) calculated the distance modulus of 
several individual stars, finding that some of them gave values near to 
$V_0 - M_V = 11.15-11.35$, while other early-type stars gave somewhat 
smaller values ($V_0 - M_V = 10.5-11$).
A more robust determination by Prisinzano et al. (2005), considering a 
sample that reaches down to $V\sim23$, lead to the value of 
$V_0 - M_V = 10.48$, which means a significantly lower cluster distance 
of 1.25~Kpc. 
This distance determination is based on more complete colour-magnitude 
diagrams, for which it is possible to define the blue envelope of 
the star distribution with $V>14$ because of the obscuration effect
of the giant molecular cloud which strongly reduces the detection
of background stars at optical wavelengths.

Although this indetermination in the distance to M8 does not change the main
conclusions of this work, we explore the possibility of deriving our own
estimate by considering a group of stars around Her\,36 with known spectral 
types.
We gathered optical photometry from Sung et al. (2000) and from the 
Galactic O Star Catalog (Ma\'{\i}z-Apell\'aniz et al. 2004a), as well as
near-infrared photometry from this work or from the 2MASS All-Sky
Point Source Catalog (PSC) (Second Incremental Data Release, Cutri et al.
2000) for Her\,36, 9\,Sgr (HD\,164794), W\,9 (HD\,164816), SCB\,182,
SCB\,325 and SCB\,354 (hereafter we use ``SCB~number'' to refer to the sources
from Sung et al. 2000). 
For the O-type stars, the adopted spectral types were either obtained from 
Ma\'{\i}z-Apell\'aniz et al. (2004a) (Her\,36, 9\,Sgr) or derived from our 
spectroscopic data (W\,9).
The stars 9\,Sgr and W\,9 are both suspected to be spectroscopic
binaries (cf. Mason et al. 1998).
Arias et al. (2005) have detected both components in W\,9, 
resulting a spectral classification of  O\,9.5V + B0\,V, 
and have also confirmed the binary nature of 9\,Sgr and Her\,36.
For B-type stars (SCB\,182, SCB\,325, SCB\,354) we adopted the spectral types 
from van den Ancker et al. (1997).

\begin{figure}
\includegraphics[width=85mm, bb= 28 28 566 541]{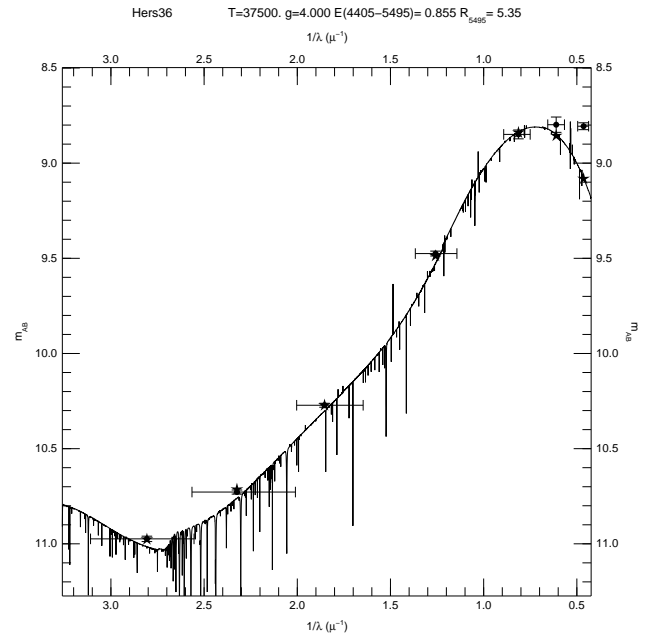}
\caption{$UBVIJHK_s$ photometry (data with error bars), SED
corresponding to the best CHORIZOS fit (spectrum), and synthetic
photometry associated with the best SED (stars) for Her\,36. For
the photometry, vertical error bars display photometric uncertainties
and horizontal ones display the approximate extent of the passband. The
$H$ and $K_s$ filters were not used for the CHORIZOS fit due to the infrared
excess of Her\,36, which is readily apparent in the plot.}
\label{ajuste}
\end{figure}

\begin{table*}
\centering
\caption{Individual distance modulii derived with CHORIZOS for some early-type stars in M\,8.}
\label{distance}
\begin{tabular}{ccccccccc}
\hline\\
Name          &  Spectral type  &   $V$    &   ${M_{V}}^\ast$ &    $ E(4405-5495)^\dag$   &    $R_{5495}^\dag$   &  $A_V$   & $V_0-M_V$ & $\chi^2$ \\
              &                 &          &                   &                      &                 &          &   & \\
\hline\\
Her\,36       &   O7.5\,V(n)    &  10.297  &  $-4.80\pm0.20$ &  $0.85\pm0.01$ & $5.39\pm0.09$  & $4.60\pm0.04$ & $10.50\pm0.20$ & 2.32 \\     
9\,Sgr        &   O4\,V((f))    &   5.966  &  $-5.50\pm0.20$ &  $0.32\pm0.02$ & $3.93\pm0.25$  & $1.26\pm0.03$ & $10.21\pm0.20$ & 1.00 \\  
W9            &   O9.5\,V+B0\,V &   7.089  &  $-4.60\pm0.20$ &  $0.28\pm0.03$ & $3.33\pm0.25$  & $0.94\pm0.07$ & $10.75\pm0.21$ & 2.34 \\    
SCB\,182      &   B2\,V         &   9.953  &  $-2.20\pm0.20$ &  $0.39\pm0.02$ & $4.50\pm0.20$  & $1.77\pm0.04$ & $10.52\pm0.20$ & 0.90 \\  
SCB\,325      &   B7\,V         &  11.552  &  $-0.60\pm0.20$ &  $0.27\pm0.02$ & $4.09\pm0.26$  & $1.11\pm0.04$ & $11.04\pm0.20$ & 2.75 \\   
SCB\,354      &   B7\,V         &  11.922  &  $-0.60\pm0.20$ &  $0.25\pm0.01$ & $3.79\pm0.20$  & $0.97\pm0.03$ & $11.55\pm0.20$ & 2.11  \\    
              &                 &          &                 &                &                &     & \\             
\hline\\                        
\multicolumn{8}{l}{\footnotesize $^\ast$ Values from Walborn (1972, 1973).}\\
\multicolumn{8}{l}{\footnotesize $^\dag$ $E(4405-5495)$ and $R_{5495}$ are the monochromatic
equivalents to $E(B-V)$ and $R_V$, see Ma\'{\i}z-Apell\'aniz (2004).}\\
\end{tabular}
\end{table*}


We used the code CHORIZOS developed by Ma\'{\i}z-Apell\'aniz (2004)
to derive the individual distance modulii of these stars. 
CHORIZOS is a code that uses $\chi^2$ minimization to find all models 
compatible with the observed
data set in the model ($N$-dimensional) parameter space, in our case,
broadband photometry and spectral types. 
For a complete description of how this method works 
see Ma\'{\i}z-Apell\'aniz (2004). 
We considered TLUSTY (Lanz \& Hubeny 2003) atmosphere models  
and Kurucz (Kurucz 2004) models 
for the spectral energy distribution (SED) of O-type and B-type stars,
respectively. 
Figure~\ref{ajuste} shows the 
SED corresponding to the best CHORIZOS fit (spectrum) and the synthetic
photometry associated with the best SED (stars) for Her\,36.
The photometric data used for the CHORIZOS fit are also shown in the figure. 
$H$ and $K_s$ filters were not considered due to the infrared excess of 
Her\,36, which is readily apparent in the plot.
Table~\ref{distance} lists all the compiled quantities and the derived
parameters, i.e., the colour excess $E(4405-5495)$ and the ratio of total 
to selective absorption $R_{5495}$, which are the monocromatic equivalents 
to the usual $E(B-V)$ and $R_V$, respectively (for a detailed explanation
see Ma\'{\i}z-Apell\'aniz 2004), and the distance modulus. 
The $\chi^2$ values corresponding to the best fit in each case have also been 
included in the last column.
The main source of error comes from the adopted values of the absolute 
magnitudes. 
The $R_{5495}$ value is close to the canonical one of 3.1 only for 
W9; for the rest of the stars it shows "anomalous" values.
In particular, the value of $R_{5495}=5.36$ derived for Her\,36, which 
represent one of the highest $R_V$ values known 
(cf. Cardelli, Clayton, \& Mathis 1989),
is in excellent agreement with earlier estimations.
Besides, as mentioned before, the analysis of its SED using CHORIZOS shows
that this star presents an evident infrared excess, which is also apparent
in the near-infrared colour-colour diagram (see Section~\ref{los_diagramas}).

\begin{figure}
\includegraphics[width=85mm]{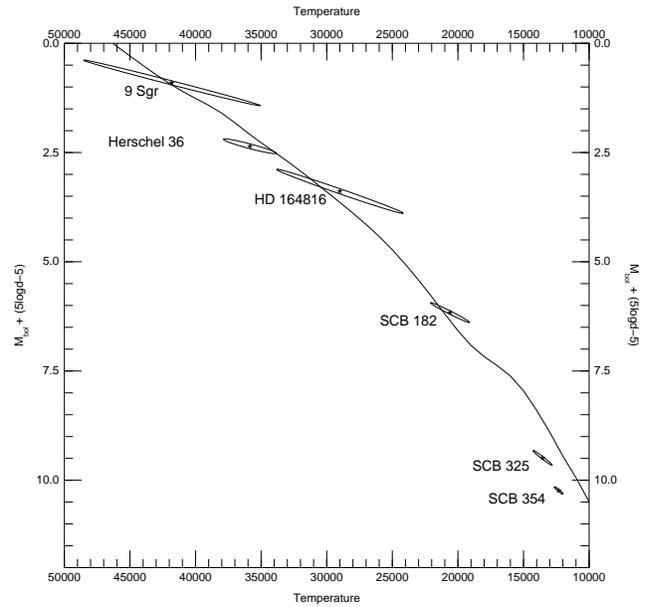}
\caption{HR diagram compiled from the CHORIZOS output for the six
stars in our sample. Ellipses indicate the 68\% likelihood contour for
each star. The line shows the main sequence for $5\log d - 5 = 10.5$.
As seen in the plot, such a distance is compatible with the CHORIZOS
output for the four earliest stars but not for the latest two.}
\label{diagHR}
\end{figure}

As also shown in Table~\ref{distance}, the four earliest stars lead to 
an average distance modulus of nearly 10.5 (1.25~Kpc), clearly lower than the 
previously adopted by other authors and similar to the value derived by 
Kilambi et al. (1977). 
On the other hand, the distance modulii derived for the two B7-type stars
is somewhat larger ($\sim11.3$).
Such problem is illustrated in the Hertzprung-Russell diagram compiled 
from the CHORIZOS output 
which is shown in Figure~\ref{diagHR}.
Ellipses indicate the 68\% likelihood contour for each star.  
The main sequence for $5\log d - 5 = 10.5$ has been marked with a solid line.
As seen in the plot, such a distance is compatible with the CHORIZOS
output for the four earliest stars but not for the latest two.

We note here that the association of 9\,Sgr, located $\sim3'$ to the NE of
Her\,36, with the Hourglass region is based not only in a common distance 
but also in the observed morphology of the gas and dust in the region. 
Besides a dust ``finger-shape'' structure that will be discussed later, 
the northern part of the nebula seems to be illuminated by this massive star.
This is especially evident in the photodissociation
regions (PDRs) facing to 9\,Sgr observed in {\em HST} images (see 
Figures~\ref{s2_map} and \ref{s2ha_map} in Section~\ref{signatures}).

Based on the values derived for the earliest stars, which are systematically
smaller than the previously published, we adopt
a distance modulus of $V_0 - M_V = 10.5$ for the Hourglass region.
A posible explanation for the disagreement with the values 
derived by other authors is considered here.

As pointed out by van den Ancker et al. (1997), the study of a very young
open cluster like NGC~6530 can not be performed using average extinction
laws, since each star has its own individual extinction characteristics. 
In fact, anomalous extinction laws have been found for several stars embedded 
in the cluster. This may be a serious obstacle when deriving distance modulii 
from the observed optical colours, as done in most previous studies.
In the infrared however, the greatly reduced extinction and the independence
of the extinction law with the ratio of total to selective absorption $R_V$
(Jones \& Hyland 1980; Cardelli et al. 1989; Martin \& Whittet 1990;
Whittet et al. 1993), make distance determinations more reliable.

An alternative interpretation for the disagreement between the distance
values derived by different authors may be that, when looking at the
direction of NGC~6530, we are actually seeing distinct groups of stars
placed at different depths.
Figure\,10 in van den Ancker et al (1997) shows the distribution of the 
distances of individual stars in the cluster, from which they estimated an 
average distance of 1.8~kpc. However, the distribution has a 
remarkable secondary peak around 1.5~kpc. In this way, the two B7-type stars 
analyzed with CHORIZOS (SCB\,325 and SCB\,354) lead to distance modulii of 
11.04 and 11.55, respectively. Both stars are located to the east of the
Hourglass Nebula, in a region with lower reddening, as inferred from their
colour excesses $E(4405-5495)\sim0.26$. 
So they presumably locate deeper in the spiral arm but are being observed
through a window relatively free of dust.
Finally, as mentioned above, Sung et al. (2000) also obtained
two values for the distance modulus, but they left out the shorter distance
estimate with the argument that many of the B-type stars 
were probably binaries.

Distance estimations to such very young open clusters are extremely
complicated since they require the knowledge of the extinction, which is really
due to circumstellar rather than intracluster material
(van den Ancker et al. 1997). Each star have its own extinction law, and
hence a very detailed study is needed.
We conclude that a definitive distance to the Hourglass Nebula is not yet 
established, but the CHORIZOS analysis of the earliest stars in the region 
is in favour of a distance modulus of 10.5 (1.25~kpc).

\subsection{Photometry of infrared sources}

Point-spread function (PSF) photometry was performed using IRAF/DAOPHOT-II 
software running in a Linux workstation. Stellar-like sources with fluxes 
significantly above the mean background were extracted 
using a threshold flux of $3\sigma$. 
A set of PSF star candidates was selected in each image avoiding 
nebular and crowded regions. 
The PSF was calculated using a {\it lorentz} function and one look-up table. 
The final PSF photometry was performed on all the extracted sources
using an aperture 3 pixels in radius. This corresponds to an aperture
diameter which is roughly that of the measured full width at half-power 
of a typical stellar image.
Aperture corrections were computed between apertures of 3 and
15 pixels in radius using the same PSF stars.  
The final number of detected sources was 762 in $K_s$, 748 in $H$ and 653 in 
$J$ with sensitivity limits of 17.5, 18.5 and 19.2, respectively. 
The photometric limits were assumed to be the magnitude corresponding to
the peak of the star distribution. This maximum occurred at about 1.5 mag 
brighter than the magnitude where the distribution fell to zero, i.e., 
16.0 in $K_s$, 17.0 in $H$ and 17.7 in $J$.
The average photometric errors in all colours are 0.03 for sources 
with mag brighter than these limits and up to 0.2 for the faintest sources. 
Table~4 (in separated file) gives the position and photometry of all the 
infrared sources detected in the field. 
Running number sources are in column~1, column~2 and 3 are right 
ascension and declination (J2000), columns~4, 5, 6, 7, 8, 9 are $J$, $H$, 
$K_s$ magnitudes and their errors, columns~10 and 11 are $J-H$ and $H-Ks$ 
colours, and column~12 contains comments related to the identification 
of optical counterparts. 

In order to detect systematic photometric errors produced by aperture
corrections and standard zero-points, we compared the magnitudes
derived from our photometry for some relatively bright stars 
with those from the 2MASS All-Sky Point Source Catalog (PSC) 
(Second Incremental Data Release, Cutri et al. 2000).
Considering only the sources in 2MASS with $photometric$ quality ``$AAA$'', 
we found an average difference in $J$, $H$ and $K_s$ magnitudes 
of $0.005\pm0.085$, $0.016\pm0.052$ and $0.008\pm0.067$, respectively, 
so no further correction was performed.

Positions in equatorial coordinates for the individual sources 
were established based on the
2MASS Point Source Catalogue (All-Sky 2003). The rms residuals between 
the positional tables from this work and from 2MASS database
are found to be 0\farcs11 and 0\farcs12, 
in $\alpha$ and $\delta$ respectively.

\subsection {Colour-Colour and Colour-Magnitude Diagrams}
\label{los_diagramas}

\begin{figure*}
\includegraphics[width=180mm]{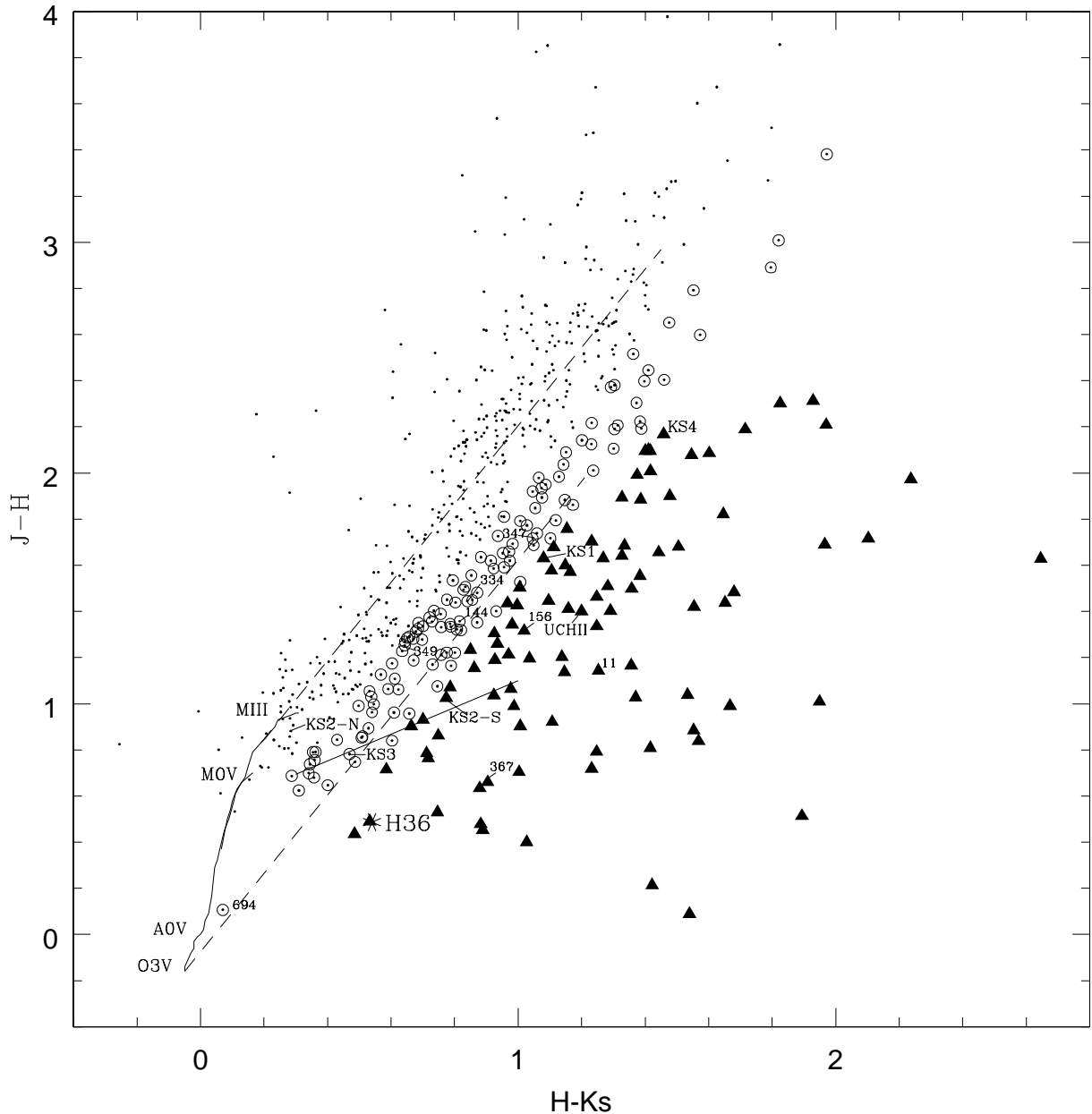}
\caption{Near-Infrared Colour-Colour diagram for the Hourglass region. The
continuous line marks the locus of main-sequence and giant stars. The two
parallel dashed lines represent the reddening vectors for a visual extinction
of $A_V = 20$ mag. The loci of classical T~Tauri stars (CTTS) from Meyer et
al. (1997) is also shown as a straight line from $H-K_s =0.3$ to $H-K_s =1.0$. 
Sources with evident infrared excesses are plotted with filled triangles. 
Open circles denote either CTTS with moderate to large reddening or
extremely reddened early-type main-sequence dwarfs. Sources marked with
points are expected to be foreground/background field objects.
Some interesting infrared sources (see text) are labelled.}
\label{diagrama_CC}
\end{figure*}

\begin{figure*}
\includegraphics[width=180mm]{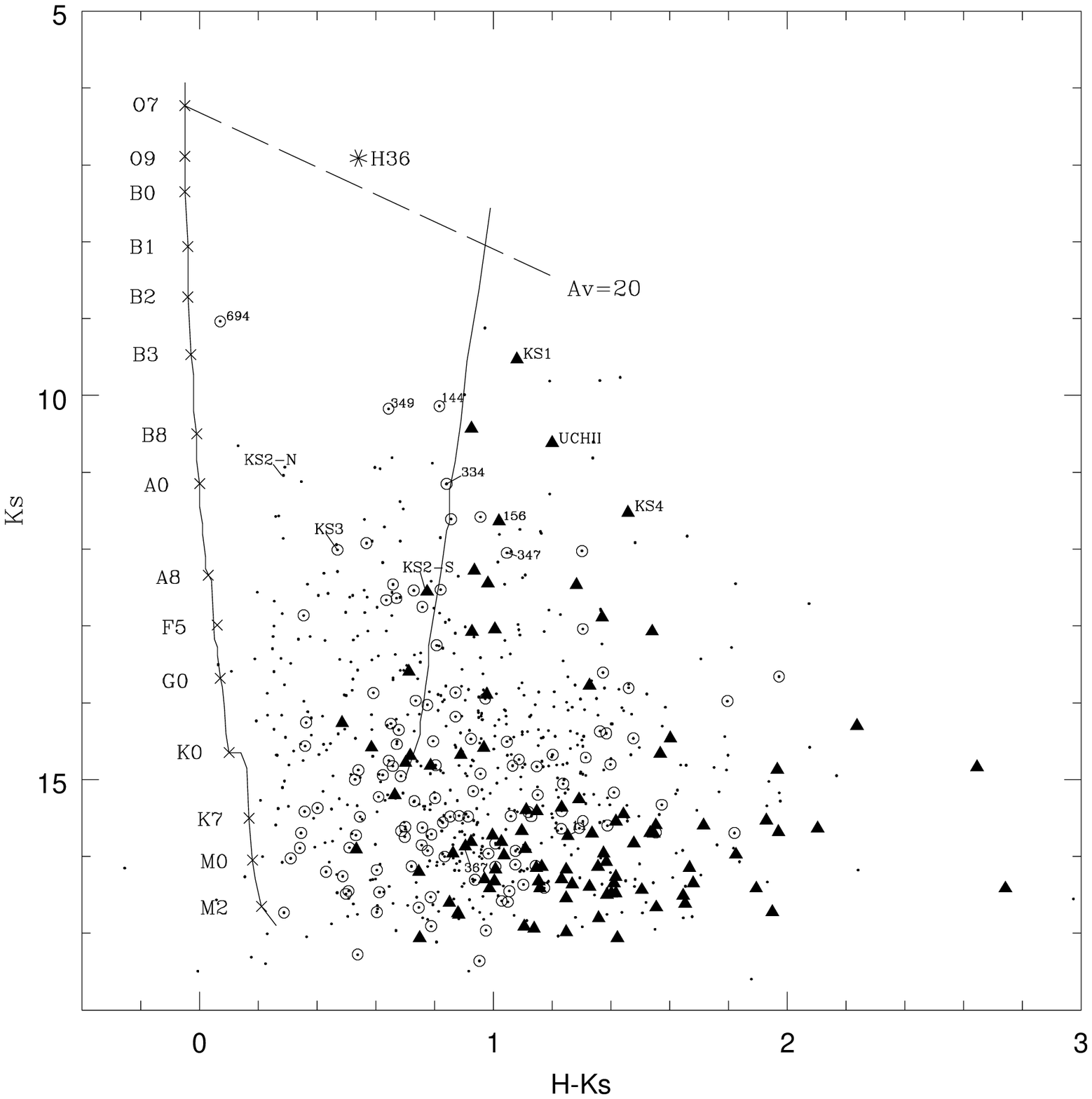}
\caption{Near-Infrared Colour-Magnitude diagram for the Hourglass region.
The location of the main sequence is shown using a distance modulus of 
10.5 mag. The red giants are represented by a vertical solid line, 
shifted to the distance of the Galactic Bulge (8~kpc) and reddened by 
$E(H-K_s)=0.64$ because of the interstellar component.
The dashed line represents the standard reddening vector with length 
$A_V=20$~mag. Symbols and object labels are the same as in 
Figure~\ref{diagrama_CC}.}
\label{diagrama_CM}
\end{figure*}

A near-infrared colour-colour (CC) diagram of the Hourglass region 
is illustrated in Figure~\ref{diagrama_CC} for the 647 sources 
in the field detected in all three wavelength bands. 
Also plotted in Figure~\ref{diagrama_CC} is the locus of points
corresponding to the position of the unreddened main sequence and
the position of the red giants.
The two parallel dashed lines represent the reddening vectors for early-
(O3\,V) and late-type (M0\,III) stars, 
determined using the values of extinction from Rieke \& Lebofsky (1985);
their length corresponds to $A_V = 20$ mag.
Figure~\ref{diagrama_CM} shows the $K_s$, $(H-K_s)$ colour-magnitude 
(CM) diagram for the same stars.
The position of the main sequence 
has been plotted, corrected to an apparent distance modulus of 10.5,
which, as was previously discussed, is appropiated for the 
Hourglass region.
The reddening vector for an O7\,V is also plotted for a visual extinction 
of 20 mag. 
The vertical solid line denotes the location of giant stars (Koornneef 1983;
Zombeck 1990) for the distance of the Galactic Bulge (8~kpc) and reddened by 
$E(H-K_s)=0.64$ due to the interstellar component.
The position of Her\,36, using near-infrared data from 2MASS, 
has been marked for reference.
We note here that, if we compare the data with the zero-age main sequence
(ZAMS) instead of the main sequence, both Her\,36 and SCB\,182 
(source \#694) fall about 1 mag 
above the expected location according to their spectral types
(O7.5\,V and B2.5\,V, respectively). This shift is even larger ($\sim$\,2 mag)
considering the distance modulus of 11.3 suggested by other authors.

\begin{figure}
\includegraphics[width=85mm]{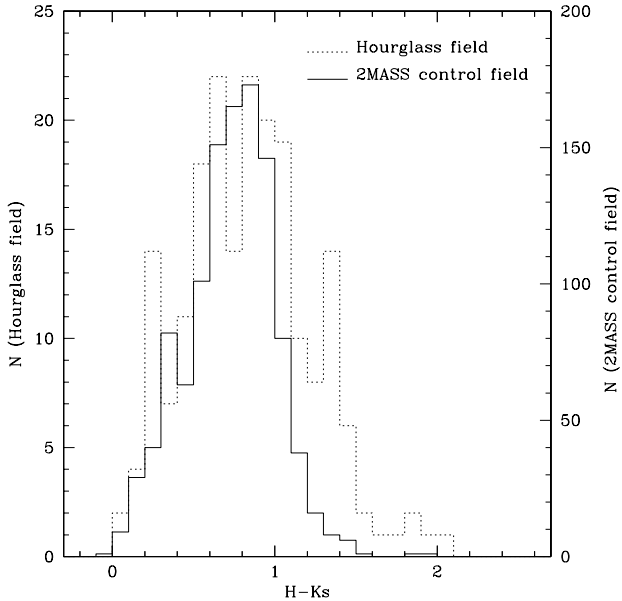}
\caption{Distribution of $(H-Ks)$ colours for the Hourglass field (dashed line)
and the 2MASS control field (solid line). The peak around $(H-Ks)=1.4$ 
observed in the Hourglass field can be account by the near-infrared-excess 
sources associated with the cloud.}
\label{histogramas}
\end{figure}

As M8 is located not far from the direction to the Galactic Centre,
many field interlopers can be expected. In particular, it is very
probable to be intercepting part of the giant population of
the galactic inner disk and Bulge.
In this way, the CC and CM diagrams show a superposition of 
well-distinguished stellar components. 
Unfortunately, no control fields were taken for these observations,
but in order to account for the distribution of the
foreground/background field stars, $JHK_s$ photometry corresponding to 
a 5 arcmin radius circular field located off the nebula 
at $\alpha(2000)=270.654$ and $\delta(2000)=-24.618$ 
($\sim$\,18' in $l$ at the same $b$) was retrieved 
from the 2MASS catalog. We considered only the sources with photometric 
errors lower than 0.1 mag.
This field was chosen to be free of significant interstellar material 
by examination of IRAS surveys in the region. 
Based on these data we studied the $(H-K_s)$ distribution of the field 
stars, which shows a significant concentration of sources around 
$(H-K_s)=0.8$, probably indicating the average colour of the Bulge 
red giants which must be highly affected by interestellar reddening
(see Figure~\ref{histogramas}).
A similar analysis for the sources with $K_{s}<14$ of the Hourglass field 
revealed a peak 
around $(H-K_s)=1.0$, with a considerable spread to larger values, 
due to the differential extincion originated in the molecular cloud. 
In addition, both histograms show a secondary peak around 
$(H-K_s)=0.3$, probably associated with 
main sequence stars moderately affected by reddening.
We note here that the number of infrared sources detected in the Hourglass 
field almost doubles the number expected from the ratio of the areas of the on
and off field observations. Two facts can account for this: the 
existence of an infrared cluster around Her\,36 and the much higher 
resolution of our observations. 

\begin{figure*}
\includegraphics[width=180mm, 
]{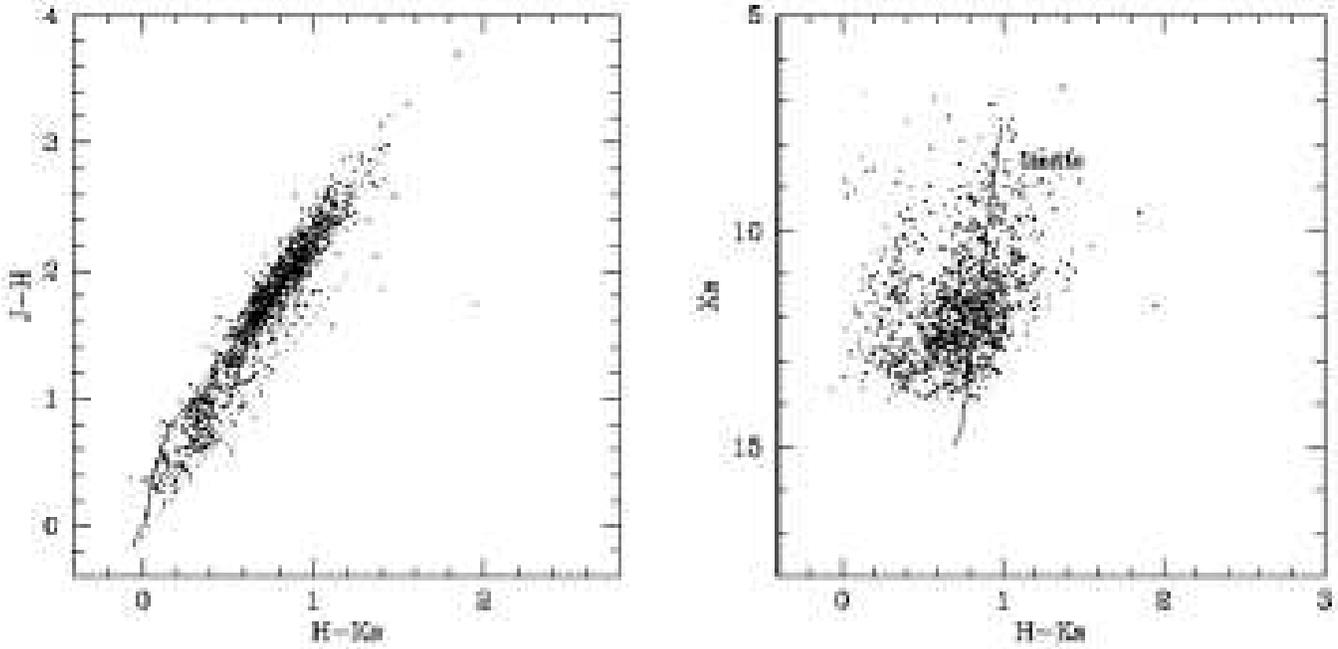}
\caption{Near-Infrared Colour-Colour and Colour-Magnitude diagrams for the 
2MASS control field. The solid line represent the location of giants for the
distance of the Galactic Bulge (8~kpc) and reddened by $E(H-K_s)=0.64$ mag
because of the interstellar component.}
\label{control_diag}
\end{figure*}

Using 2MASS data we also constructed the CC and CM diagrams for the control 
field (Figure~\ref{control_diag}). The CC diagrams for the control field and 
the Hourglass region clearly differ. The stars in the control field appear 
confined in two regions: a very crowded region corresponding to the highly 
reddened giants of the Bulge and disk, and a less populated region associated 
with field main sequence stars with little reddening. On the other hand, the 
stars in the Hourglass region spread over a much larger area of the CC 
diagram. Besides the two components observed in the control field, which 
confirm that a significant fraction of the stars are background field stars 
unrelated to the cloud, there is a significant number of sources 
located along and to the right of the reddening vector of an O dwarf star.
These stars with intrinsic colours indicative of near-infrared excess 
emission must be associated with young stellar objects (YSOs), 
such as Class~I ``protostars'', Herbig Ae/Be objects and T Tauri stars.

In Figure~\ref{diagrama_CC} two kinds of symbols have been used in order 
to discriminate between objects of different nature. In this way, filled 
triangles represent sources with evident infrared excess, which are prime 
candidates to be included among the YSOs previously mentioned. 
On the other hand, open circles may denote either classical T~Tauri stars 
affected by
moderate to large amounts of reddening, or extremely reddened early-type 
main-sequence stars. For example, according to its position in the CC and CM 
diagrams, source \#349 could be either a classical
T~Tauri star with $A_V\sim\,4$ 
mag, or an early B-type star affected by more than 11 mag of extinction.
Spectroscopic observations are needed for a definitive conclusion.

Also evident from Figure~\ref{diagrama_CC} is the lack of stars within the 
lower part of the reddening band $[0<(H-K_s)<0.3]$, reflecting a threshold 
in the extinction caused by the presence of the molecular cloud.
Just beside this gap, there is a group of sources vertically distributed
around $(H-K_s)=0.3$, which presumably represents the main sequence at the
Hourglass distance. This idea is supported by the presence of a similar
distribution $(H-K_s)=0.3$ to the right and parallel to the main sequence 
of the CM diagram. If so, the mean extinction toward the region can be 
estimated from a simple approximation (Rieke \& Lebobfky 1985), 
$A_K=1.78 \times E(H-K)$,
adopting for the main sequence stars an average intrinsic colour excess of 
0.2 mag. This leads to $A_K=0.36$ mag ($A_V=A_K/0.112=3.2$ mag).

The CM diagrams for the control and the Hourglass fields are also clearly
different. The bulk of the stars in the control field distribute in a band 
around $(H-K_s)=0.8$ (which approximately coincides with the vertical line
that represents the location of red giants for the distance of the Galactic 
Bulge), in contrast with the Hourglass field, 
where the sources appear in a wider range of $(H-K_s)$ and a great number 
of them show large values of this colour.
The large $(H-K)$ colour for these sources is likely intrinsic and due to
excess infrared emission, as deduced from their locations 
in the infrared excess region of the CC diagram.

Before concluding this section, we want to stress that the comparison of the 
infrared photometric diagrams suggests that an important fraction of the 
stars observed toward the 
Hourglass Nebula are background field stars unrelated to the cloud, 
presumably red giants of the galactic inner disk and Bulge. 
These objects are significantly reddened as a result of interestellar
extinction through the line-of-sight, plus a 
diferential contribution of an inhomogeneus molecular cloud, which, 
as will be discussed in next section, posseses clumps dense enough to block 
background radiation.

\begin{figure*}
\begin{minipage}{130mm}
\includegraphics[width=110mm]{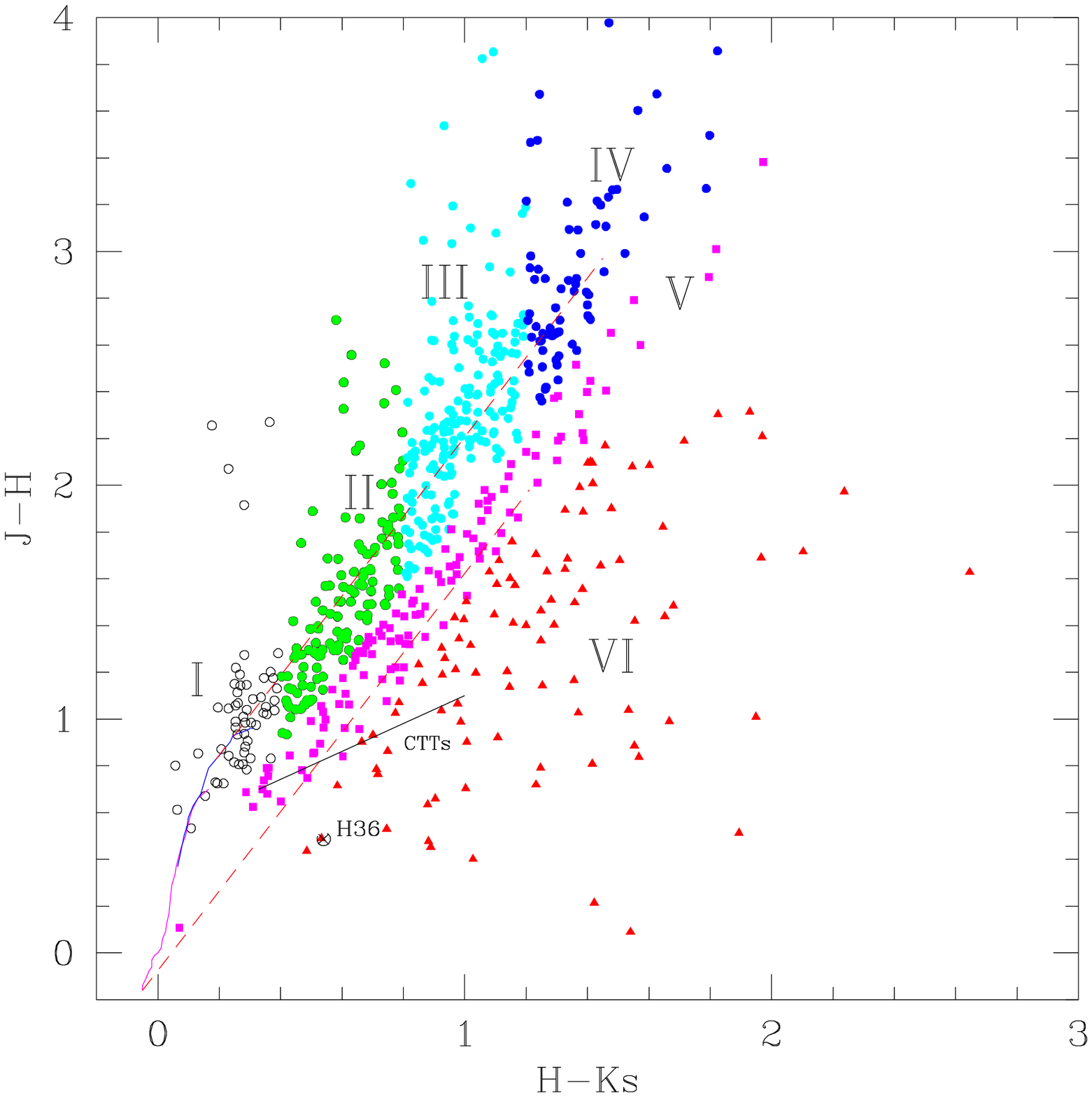}
\includegraphics[width=110mm]{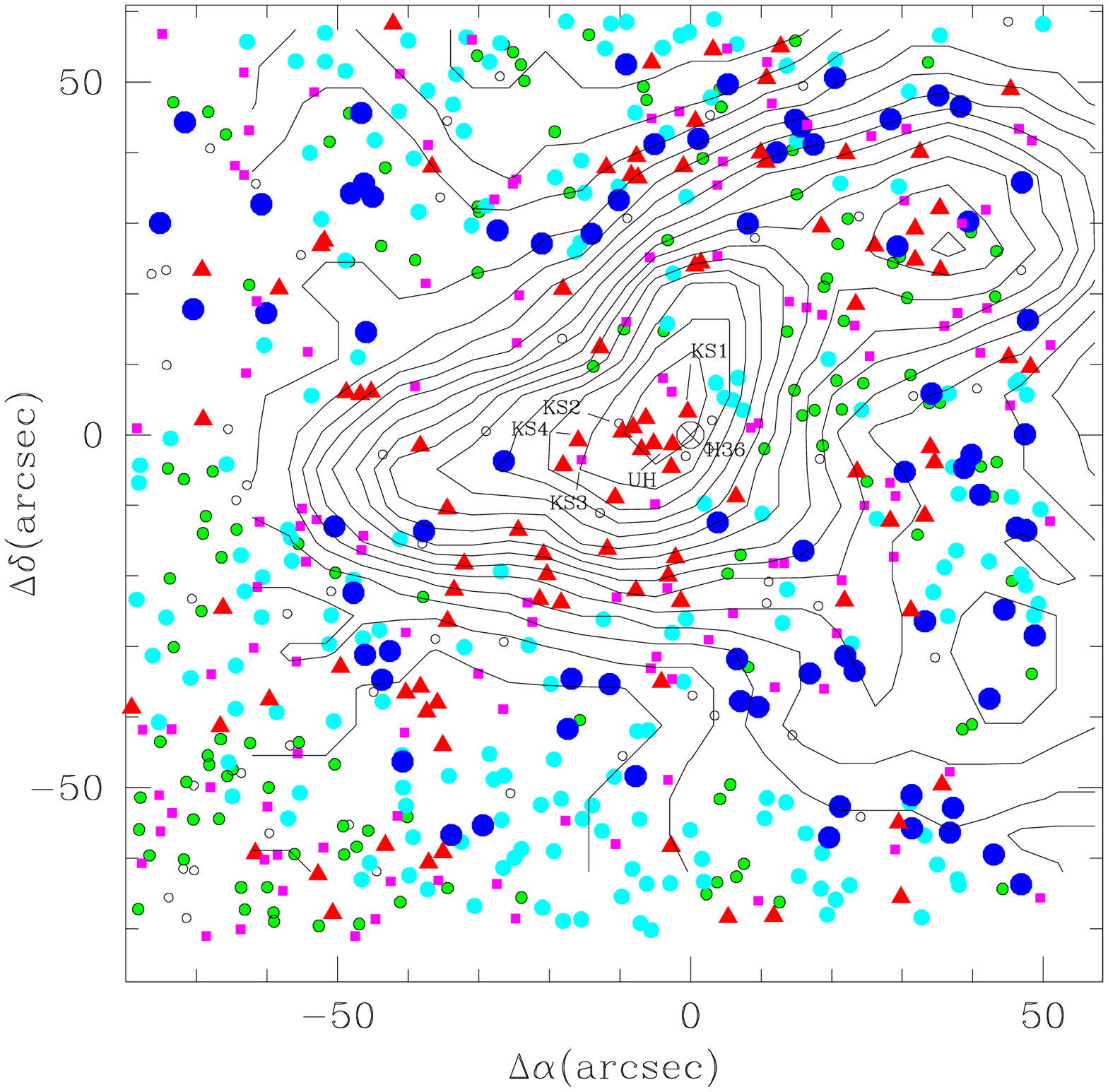}
\end{minipage}
\caption{Upper panel: CC diagram for all the sources in the Hourglass 
Nebula. Open, green, cyan and blue circles denote sources in groups I, II, III
and IV, respectively (referred as ``giants'' in the text). 
Magenta squares represent the sources in group V, 
and red triangles the genuine IR-excess objects of group VI.
Lower panel: Spatial distribution of the different stellar
groups observed towards the Hourglass Nebula. The positions of Her\,36,
the infrared sources KS\,1 to KS\,4 and the ultracompact H\,{\sc ii} 
region G5.97-1.17 are marked.   
CO $J=3-2$ contours from White et al. (1997) are also shown for comparison.
Symbols are the same as in the previous panel.}
\label{colores}                                               
\end{figure*}

As previously mentioned, when we observe in the direction of M8 we are
actually seeing a superposition of distinct stellar populations placed at 
different distances from us, which significantly complicates the interpretation
of the results. 
Trying to clarify this situation, we divided the infrared sources observed 
toward the Hourglass in six groups according to their infrared colours.
Groups I, II, III and IV include sources located above the track  
($J-H = 1.7 * (H-K_s) + 0.2$), with $H-K_s<0.4$, $0.4<H-K_s<0.8$, 
$0.8<H-K_s<1.2$ and $H-K_s>1.2$, respectively. 
This track has been chosen as the straight line parallel to the reddening 
vectors, passing approximately 
through the blue end of the classical T~Tauri loci of Meyer et al. (1997).
In some way it may be considered as a dividing line between early- and 
late-type stars, as it coincides with the reddening track of a mid-F type 
dwarf. Since these sources are
expected to be members of the stellar population of the inner disk and Bulge,
we call them ``giants'', even though group I may have a contribution of 
late-type main sequence dwarfs with lower reddening. 
Groups V and VI include sources below the former track.
Whereas group V is expected to contain both classical T~Tauri stars
with lower reddening and/or highly reddened early-type dwarfs, 
group VI will include
only objects with genuine infrared excess.
In Figure~\ref{colores} we plotted the spatial distribution of each stellar 
group using different symbols and colours. A colour CC diagram has been 
included for reference. 

Furthermore, in order to quantify the complex distribution of sources in the 
region, we subdivided the observed field in nine areas of about 
45''$\times$45'' (which we call from upper left to lower right, NE, N, NW, E, 
C, W, SE, S and SW, respectively), and counted the number of stars of each
group in every subregion.
The resulting star counts are presented in 
Table~\ref{conteos}. The interpretation of Figure~\ref{colores} and 
Table~\ref{conteos} will be discussed in the following subsections.

\begin{table*}
\centering
\caption{Star counts of different stellar populations in the Hourglass region.}
\label{conteos}
\begin{tabular}{ccccccccccc}
\hline\\
Subregion      &  NE    &   N    &     NW    &     E    &     C    &     W    &    SE   &      S    &    SW   &    sum  \\
Group          &        &        &           &          &          &          &         &           &         &          \\
\hline\\
I              &  6     &    5   &      5    &     9    &     7    &     6    &    12   &      5    &     3   &     58\\     
II             & 10     &   17   &     15    &    12    &     5    &    19    &    33   &      9    &     7   &    127\\  
III            & 14     &   29   &      8    &    20    &     9    &    23    &    22   &     38    &    17   &    180\\    
IV             &  9     &    8   &     13    &     4    &     2    &    11    &     4   &      8    &    14   &    73 \\  
V (ms)         & 10     &   13   &     15    &    12    &    10    &    16    &    25   &     11    &     7   &    119\\   
VI (ex)        &  6     &   11   &     14    &     6    &    27    &     9    &    13   &      5    &     4   &    95\\    
               &        &        &           &          &          &          &         &           &         &        \\             
sum            & 55     &   83   &     70    &    63    &    60    &    84    &    109  &     76    &    52   &     652\\  
\\
\hline\\                                                                                                              
``giants''     & 39     &   59   &     41    &    45    &    23    &    59    &    71   &     60    &    41   &     438\\  
(I+II+III+IV)  &        &        &           &          &          &          &         &           &         &        \\
ms+ex          & 16     &   24   &     29    &    18    &    37    &    25    &    38   &     16    &    11   &     214\\  
(V+VI)         &        &        &           &          &          &          &         &           &         &        \\
\\
\hline\\
ratio ex/giants     &  0.15  &  0.19  &    0.34   &   0.13   &   1.17   &   0.15   &   0.18  &    0.08   &   0.10  &    0.22\\  
                    &        &        &           &          &          &          &         &           &         &        \\
ratio (ms+ex)/giants&  0.41  &  0.41  &    0.71   &   0.40   &   1.61   &   0.42   &   0.54  &    0.27   &   0.27  &    0.49\\
                    &        &        &           &          &          &          &         &           &         &        \\
ratio ex/ms         &  0.60  &  0.85  &    0.93   &   0.50   &   2.70   &   0.56   &   0.52  &    0.45   &   0.57  &   0.80\\
                    &        &        &           &          &          &          &         &           &         &        \\
\hline\\
\end{tabular}
\end{table*}

\subsection{The extinction map}

We note here that, in contrast to other studies of dust extinction 
in molecular clouds (for example: Lada et al. 1994), it is not easy 
to infer the mean intrinsic colour $(H-K)$ of background stars 
from the control field stars, since these objects themselves suffer 
significant random extinction due to a large depth in the line-of-sight. 
Anyway, we choose  $(H-K)=0.15$ as a representative value, which is 
equivalent to the colour of an early K star, and follow the procedure 
described in Lada et al. (1994), in order to map the distribution of colour
excesses (and hence, extinctions) through the cloud. We derived the 
mean colour excess for all the sources located along the reddenning vector of 
a giant star in the Hourglass CC diagram, i.e. all the candidate late-type 
stars of the galactic inner disk and Bulge. In a rough approximation,
these can be considered as the objects in the groups I, II, III and IV 
introduced in section~\ref{los_diagramas}.
These measurements were then 
converted to a mean extinction in the $K$ band, $A_K$, using the reddening 
law from Rieke \& Lebosfky (1985):

$A_K=1.78\,E(H-K)$

\noindent We consider $A_K$ instead of the visual extinction 
$A_V=15.9\,E(H-K)$,
since the latter expression is valid only to the extent that the reddening 
law in this cloud is a normal one.  
Disregarding the objects in groups V and VI (plotted with magenta and red, 
respectively), which can be roughly adopted as belonging to the M8 region 
itself, Figure~\ref{colores} (bottom)
represents a randomly sampled map of the distribution of extinctions 
across the Hourglass nebula, obtained by plotting the position of each field
star with a circle whose size and colour 
depend on the total extinction derived from its infrared colours
(from the largest to the smallest extinction, we have big blue, cyan, green 
and open circles).
In spite of the depth of our infrared observations, the centre of the cloud 
remains devoid of stars, indicating the presence of extremely high extinctions.
Also evident is the existence of an extinction gradient extending 
from the position of Her\,36.
In Figure~\ref{colores} we overlay a contour map of the integrated 
CO $J=3-2$ emission 
from White et al. (1997). The correspondence between our extinction map 
and the distribution of molecular material is impressive,
demostrating that the extinctions derived for the individual candidates
for background giants are good tracers of the molecular cloud.

\subsection{The infrared cluster around Herschel\,36}

By photometry, 763 sources have been extracted from a
135''$\times$139'' area around the massive star Her\,36, the vast 
majority of which had not been detected by previous observations.
652 of these sources have been detected in all three wavelength bands and 
are plotted in the CC and CM diagrams in Figures~\ref{diagrama_CC},
\ref{diagrama_CM} and \ref{colores}. 
The analysis of these infrared diagrams lead us to
suggest that an important fraction of the stars observed toward the Hourglass
are background field stars, probably red giants of the inner disk and Bulge. 
Adjusting for this 
background contamination is however extremely difficult. 
Although somewhat rough, a direct alternative is to consider the extreme case 
in which all the sources in groups I, II, III and IV are foreground/background
field stars, while all the sources in groups V and VI belong to the 
cluster (see section~\ref{los_diagramas} and Figure~\ref{colores} for the 
definition of the stellar groups). 
Then the number of infrared sources expected to be related to the region 
may be about 214. 95 of them show genuine moderate and large infrared 
excesses, yielding a fraction of near-infrared excess sources of about 
45 per cent.
We should keep in mind that we have presumably overestimated the number of
objects associated to the molecular cloud, and therefore the resulting 
fraction of infrared excess sources may be only a lower limit.
Anyway, the estimated value is close to the fraction found within 
the Taurus (50~\%) (Kenyon \& Hartmann 1995), NGC\,1333 (60~\%)
(Lada et al. 1996) and S87E (40~\%) (Chen et al. 2003) star forming regions.
A strikingly important fact is what happens in the central part of the
field (subregion C in Table~\ref{conteos}), 
where the infrared excess objects reach a total of 27 or
over 70~\% of all the point sources assumed to belong to the cluster.
From the analysis of molecular hydrogen and CO emission, Burton (2002)
suggested the existence of a molecular outflow from Her\,36. This fact, along
with the relative large fraction of infrared excess sources and the
morphological evidence of jets from some of them (see 
section~\ref{signatures}), indicate the cluster population is extremely young.
From the comparison with similar star forming regions for which ages have been
estimated (see Section~5 in Lada et al. 1996), one could say that 
this cluster is probably between $1-2\times10^6$\,years, if not younger.

CO line emission toward the Hourglass Nebula was studied by White et al. 
(1997), who detected an elongated  molecular core, 30'' $\times$ 20'' in 
size and orientated SE-NW.
The superposition of CO $J=3-2$ contours on the
spatial distribution of sources in the field is shown in
Figure~\ref{colores}. The CO emission peaks at Her\,36 and 
40'' to the NW. Note that the first peak is the second most 
intense CO source observed with a single dish antenna.

As clearly seen in Figure~\ref{colores}, the infrared excess sources do not 
distribute uniformly in the region. Instead, they extend along the molecular 
core. While Her\,36 and the infrared sources KS\,1 to K\,S4 designated by 
Woodward et al. (1990) lie close to the most intense peak, another group 
of candidate YSOs coincides with the secondary peak. Additionally, a number 
of sources with infrared excess appear grouped to the South of the Hourglass,
within the cavity open by the strong wind of Her\,36.
It is remarkable that almost no source is detected to the NE of Her\,36,
either with or whitout infrared excess, 
indicating an extremely high densiy of the molecular cloud.

The infrared excess sources distribution can be quantitatively described
using the star counts in Table~\ref{conteos}. 
Three representative ratios between distinct stellar populations have been
computed. The ratio of genuine infrared excess sources to candidate
late-type objects, ex/''giants'', has a very low value 
($\sim$ 0.15 in average) in 7 of the 9 regions in which the Hourglass field 
was subdivided (see section~\ref{los_diagramas}). 
Nevertheless, it greatly exceeds its average value in the NW region
and reaches 1.17 in the central (C) region.
These two areas aproximately coincide with the secondary and primary 
CO emission peaks respectively, confirming that the youngest objects locate 
along the molecular core.

\subsection{Individual stars}
\label{individual}

Among the most peculiar objects in the field we find the infrared sources 
designated by Woodward et al. (1990) as KS\,1 to KS\,4,
as well as the ultracompact H\,{\sc ii} region G5.97-1.17 studied by
Stecklum et al. (1998).  
What follows is a brief description of the infrared properties of these 
objects.

{\em a) KS\,1 :}
This source (\#312) is $\sim$\,3\farcs3 north and $\sim$\,0\farcs3 east of 
Her\,36. Allen (1986) suggested that this star and Her\,36 may be part of 
a Trapezium-like stellar cluster, based on the similar separation distances.
KS\,1 is actually a binary star with a very red northern component, 
as pointed out first by Stecklum et al. (1995). 
The binarity of this object compelled us to perform aperture photometry on it.
According to its near-infrared colours ($K_{s}=9.53$, $J-H=1.63$, 
$H-K_s=1.08$), it is very probably
a Herbig Ae/Be object affected by a few magnitudes of visual extinction. 

{\em b) KS\,2:}
Located $\sim$\,1\farcs5 north and $\sim$\,11\farcs1 east of Her\,36,
this star is also a binary system. In our PSF photometry we
identified two well-separated (angular separation $\sim$1\farcs3) sources: 
\#385 (north) and \#382 (south). While source \#385 (KS\,2-N) is 
presumably a late B-type dwarf attenuated by $\sim$~5 mag of visual 
extinction, the infrared colours of source \#382 (KS\,2-S) are typical of a 
classical T~Tauri star ($K_{s}=12.55$, $J-H=1.03$, $H-K_s=0.77$).

{\em c) KS\,3 and KS\,4:}
These two sources are located $\sim$16\farcs8 east of Her\,36, very near 
the apex of the optical bicone (the ``waist'') of the Hourglass.
The observed infrared colours of KS\,3 (\#410) can be replicated either by a
relatively hot $\sim$ B9/A0 dwarf affected by $\sim$~8~mag of visual 
extinction, or by a cooler T~Tauri star 
($K_{s}=12.01$, $J-H=0.78$, $H-K_s=0.47$).
On the other hand, KS\,4 (\#414) is much redder than KS\,3 and shows 
significant 
near-infrared excess emission. Its position on the CC~diagram is typical of 
very young stellar objects such as Class~I protostars 
($K_{s}=11.53$, $J-H=2.17$, $H-K_s=1.46$).

{\em d) UC H\,{\sc ii} region G5.97-1.17:} 
High resolution optical, infrared and radio observations revealed that this
object, 2\farcs7 distant from Her\,36,  may be a proplyd, i.e., a young star 
surrounded by a circumstellar disk that is being photoevaporated by the nearby
hot star (Stecklum et al. 1998). Because of its close proximity to Her\,36, 
we had to perform aperture photometry on this source (\#330). 
The near-infrared 
colours obtained indicate that G5.97-1.17 likely belong to the class of 
Herbig Ae/Be objects ($K_{s}=10.62$, $J-H=1.40$, $H-K_s=1.20$).

\subsection{Distribution of dust and gas}
\label{dist}

\begin{figure*}
\includegraphics[width=180mm]{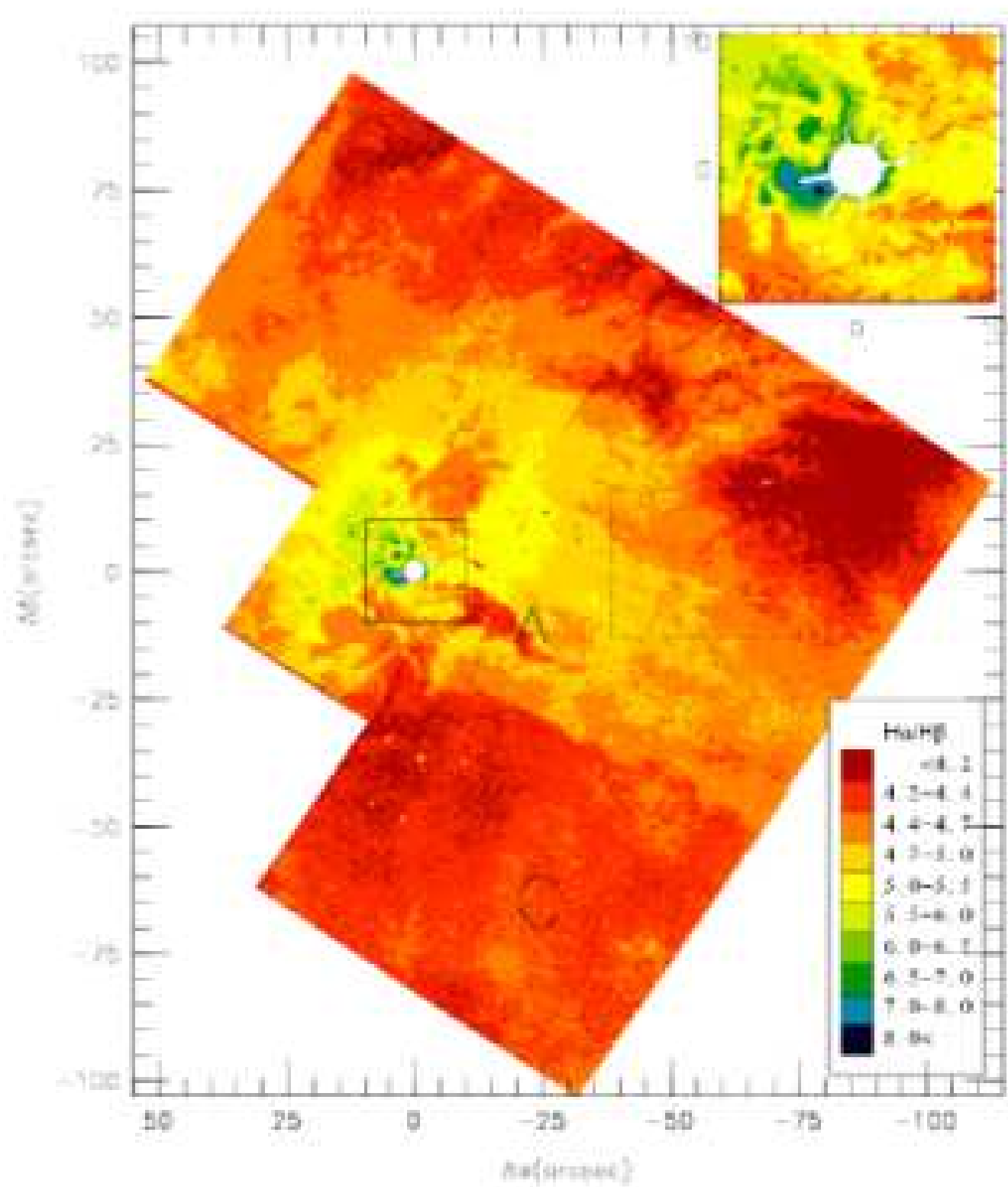}
\caption{Balmer ratio map of the Hourglass region. An 
enlargement of the central part is shown in the upper right corner of 
the panel. The letters A and C refer to the regions described in the
text (see sections~\ref{dist} and \ref{signatures}). The dashed-lined 
rectangular box defines the region shown in Fig.~\ref{hh_ring}.}
\label{Balmer_map}
\end{figure*}

\begin{figure*}
\includegraphics[width=180mm]{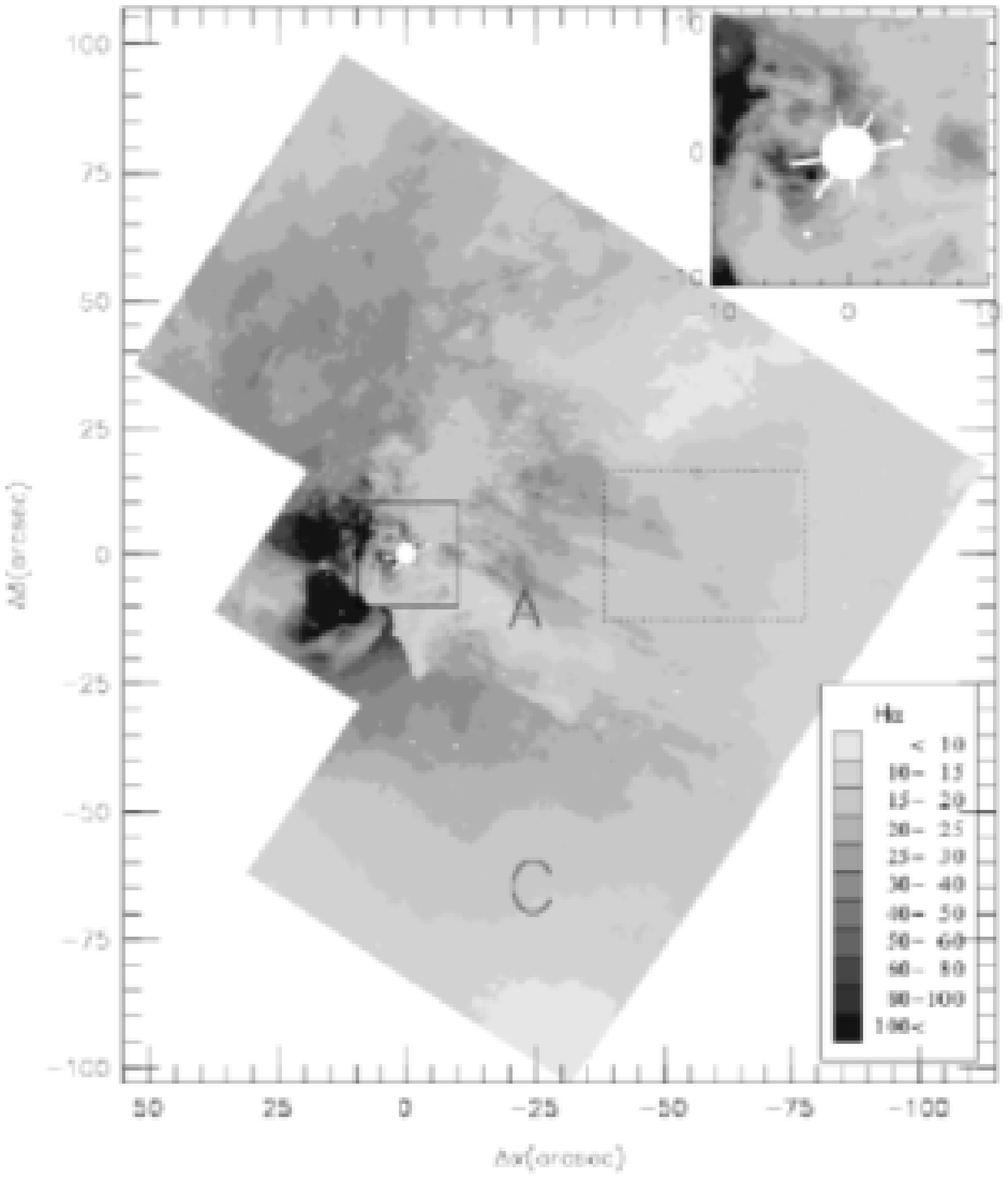}
\caption{H$\alpha$ map of the Hourglass region. An 
enlargement of the central part is shown in the upper right corner of 
the panel. The letters A and C refer to the regions described in the
text (see sections~\ref{dist} and \ref{signatures}). The dashed-lined 
rectangular box defines the region shown in Fig.~\ref{hh_ring}.
The surface brightness units are $10^{-16}$\,erg cm$^{-2}$ s$^{-1}$
per WFC pixel.}
\label{Ha_map}
\end{figure*}

Using $HST$ images, we have produced the maps of the following line ratios:
H$\alpha$/H$\beta$ (Balmer ratio) and 
([S\,{\sc ii}]6717+[S\,{\sc ii}]6731)/H$\alpha$ (S2H~ratio).
Figure~\ref{Balmer_map} shows the Balmer ratio map. The H$\alpha$ map has been
included in Figure~\ref{Ha_map} for reference. 
The Balmer ratio can be used as a tracer of extinction
by dust. For temperatures of $\sim$10$^4$\,K and electron densities of 
$\sim$100\,cm$^{-3}$,
the theoretical H$\alpha$/H$\beta$ value is 2.86 (Osterbrock 1989).
The presence of extinction increase the observed ratio, 
since it affects H$\beta$ wavelenghts more than H$\alpha$ wavelengths.
The measured ratio will depend on the geometry of the dust and gas clouds, 
as well as on the extinction law.

As seen in Figure~\ref{Balmer_map}, the Balmer ratio goes clearly above the 
expected value of 2.86 over the whole region of the Hourglass. In fact, 
it has a minimum value close to 4.0. 
We interpret this effect as consequence of the interstellar extinction 
toward the Hourglass Nebula. A colour excess $E(B-V)$ can be
associated to a measured ratio $r_B=I$(H$\alpha$)/$I$(H$\beta$) 
according to the expression:

$E(B-V)=E_0\,log_{10} (r_B/2.86)$,

\noindent where $E_0$ is a constant which depends on the reddening law. 
If we adopt the extinction law parametrized by Cardelli et al. (1989),
$E_0$ = 2.307, and then $E(B-V)$ = 0.34 when $r_B$ = 4.0.
This is in excellent agreement with the mean value of the reddening derived 
from the optical colours of the stars in NGC~6530 ($<E(B-V)>$ = 0.35; 
Sung et al. 2000). 

The distribution of dust in the central part of the Nebula is quite complex, 
showing strong variations in a scale of arcseconds, which is clearly
appreciable in the enlargement shown at the upper right corner of the panel. 
The whole central area seems to be highly reddened and the maximum values of 
$r_B$ are found around Her\,36 ($r_B>7$ and $E(B-V)>0.9$). 
The region immediately to the E of Her\,36 shows an interesting
spatial correlation between $r_B$ (Fig. 7) and H$\alpha$ (Fig. 8):
brighter (in H$\alpha$) regions present higher values of $r_B$, rather
than the opposite, which is what would be expected if optically-thin
clouds where partially occulting the ionized gas behind. Furthermore,
the correlation is also apparent when comparing the spatial
distribution of H$\alpha$ and the S2H ratio in the sense that low
values of the S2H ratio correspond to bright areas. Similar
correlations have been detected in NGC 604 (Ma\'{\i}z-Apell\'aniz et
al. 2004b) and in 30 Doradus (Walborn et al. 2002, Ma\'{\i}z Apell\'aniz et
al. 2005 in preparation) and can be explained if the variations in
extinction are caused by optically-thick clouds instead of by
optically-thin ones. In such a model, a strong variable foreground
screen is present and located between the main part of the H\,{\sc ii}
region and the observer. In those areas where the screen is thinner, we
can see into the bright part of the H\,{\sc ii} region but at the cost
of measuring large values of $r_B$ (note that values of $r_B$ larger
than about 4.0 imply that dust and gas cannot be uniformly mixed but
rather that the dust is located in a foreground screen,
Ma\'{\i}z-Apell\'aniz et al. 2004b). There, the gas we are seeing is
directly exposed to the ionizing radiation of Her\,36, located a
short distance away, and therefore shows a high excitation, implying
low values of the S2H ratio. In those areas where the screen is
thicker, dust completely blocks the bright areas of the H\,{\sc ii}
region and all we can see is the backside (as seen from Her\,36) of
the cloud, which only receives low-intensity ionizing-radiation
(scattered from Her\,36 or originating from other ionizing sources
at large distances, see Fig. 8 of Ma\'{\i}z-Apell\'aniz 2005), thus producing
H\,{\sc ii} gas with a low ionization parameter. There, the S2H ratio
is much higher, as expected, and $r_B$ is lower because there is little
dust between the source and the observer since the cloud is located at
larger distances than the source.

Extinction seems to be fairly uniform over the rest of the field.
The cavity toward the South open by the winds of Her\,36 (marked with a letter
C in Figure~\ref{Balmer_map}) suffers 
almost no obscuration, thus showing the lowest values of $r_B$, i.e.,
$4.0<r_B<4.5$, or equivalently, $0.34<E(B-V)<0.45$. 
A kind of thin flap of dust seems to cover all the
region toward the West of Her\,36. 
This fact implies that we are not actually penetrating the molecular cloud but 
only detecting the front gas H$\alpha$ emission, which can explain
the relatively low extinction measured ($r_B<5$ and $E(B-V)<0.56$).

By comparing the Balmer ratio and the H$\alpha$ emission, we find 
a number of dark regions with very low H$\alpha$ intensity (I(H$\alpha$)$<0.5$)
which also show small values of $r_B$ ($r_B\sim4.0$). 
An example of such region has been marked with a letter A in 
Figure~\ref{Balmer_map}. 
Here, a low  $r_B$ does not mean low extinction. On the contrary the optical 
depth is so high that in case of existing gas emission hidden by dust, it can 
not be detected only with optical data. Such possible emitting regions should 
be unveiled using near-infrared Br$\gamma$ imaging.

To conclude we want to stress that extinctions derived from the Balmer
emission lines are only representative of the outer ``shell'' of the nebula, 
up to about $A_V\,\sim\,3$ mag. Optical observations do not probe 
the gas and dust distribution more deeply into the cloud, and thus estimating 
ionizing fluxes and other properties only from optical data will 
lead to uncertain results. Near-infrared imaging are inevitably needed to 
analyze what processes are really taking place behind the dust.

\subsection{Signatures of ongoing star formation in {\em HST} images}
\label{signatures}

Archival {\em HST} emission-line images in H$\alpha$, [O\,{\sc iii}] and 
[S\,{\sc ii}] were searched for proplyds, jets and other features that
might be associated with the newly detected candidate YSOs.
The {\em HST} observations reveal a rich variety of structures related to star 
formation, similar to those observed in M16 and M42.

Herbig-Haro (HH) objects are shock-excited nebulae powered from YSOs
(see Reipurth \& Bally 2001). The morphological analysis of [S\,{\sc ii}]
images is a 
commonly used technique for surveying star forming regions 
for the location of excited gas arising from HH flows 
(e.g. Wang et al. 2003). This 
is helped by the analysis of [S\,{\sc ii}]/H$\alpha$ ratio maps 
in which  regions with high values indicative of shock-excited gas are 
searched.

\begin{figure*}
\includegraphics[width=180mm]{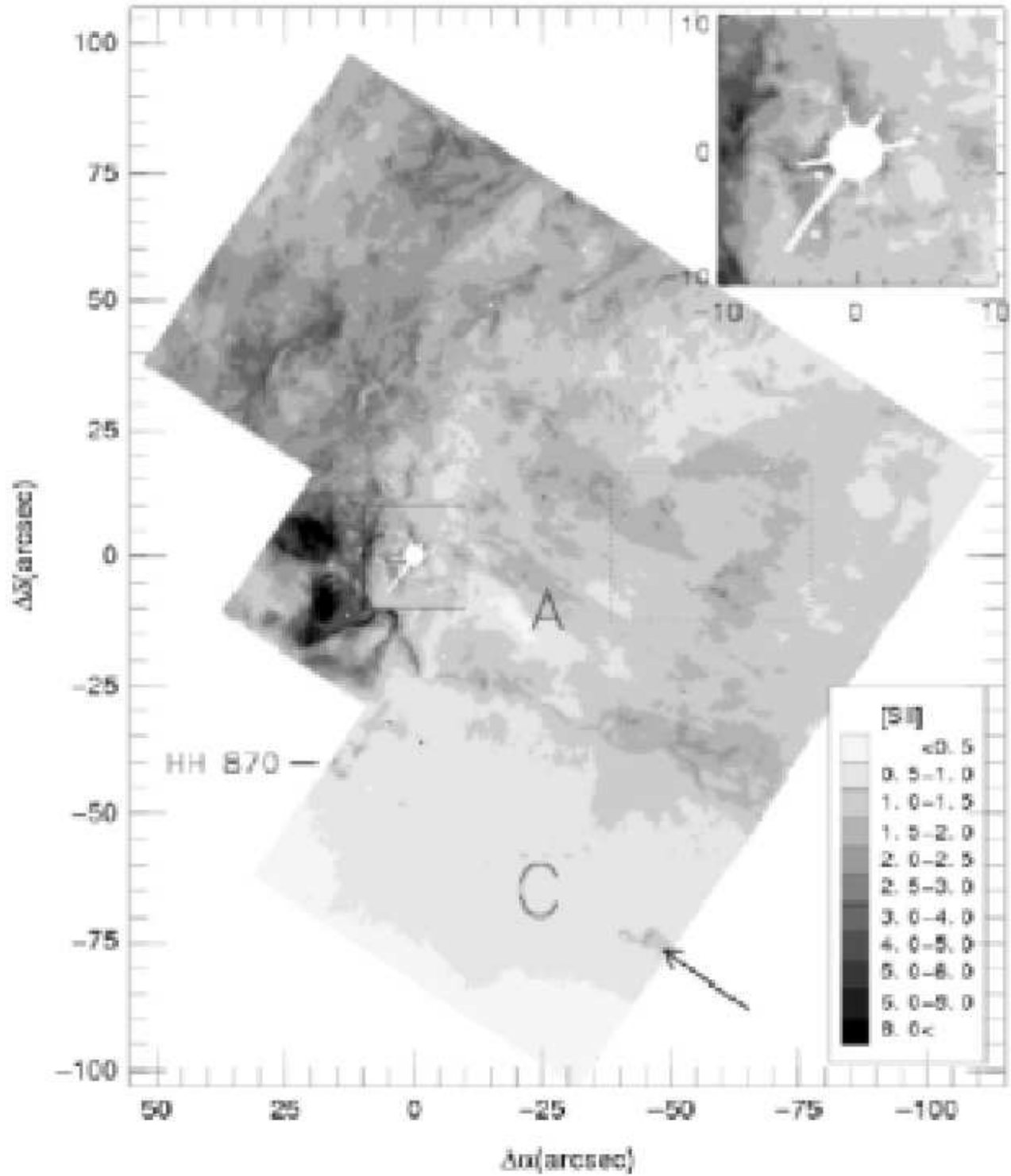}
\caption{[S\,{\sc ii}] emission map of the Hourglass region. An 
enlargement of the central part is shown in the upper right corner of 
the panel. The letters A, B and C refer to the regions described in the
text (see sections~\ref{dist} and \ref{signatures}). The dashed-lined 
rectangular box defines the region shown in Fig.~\ref{hh_ring}.
The arrow indicates the direction to the O-type star 9\,Sgr, located 
4\farcm12 from the finger-shaped molecular globule. The surface brightness 
units are $10^{-16}$\,erg cm$^{-2}$ s$^{-1}$ per WFC pixel.}
\label{s2_map}
\end{figure*}

\begin{figure*}
\includegraphics[width=180mm]{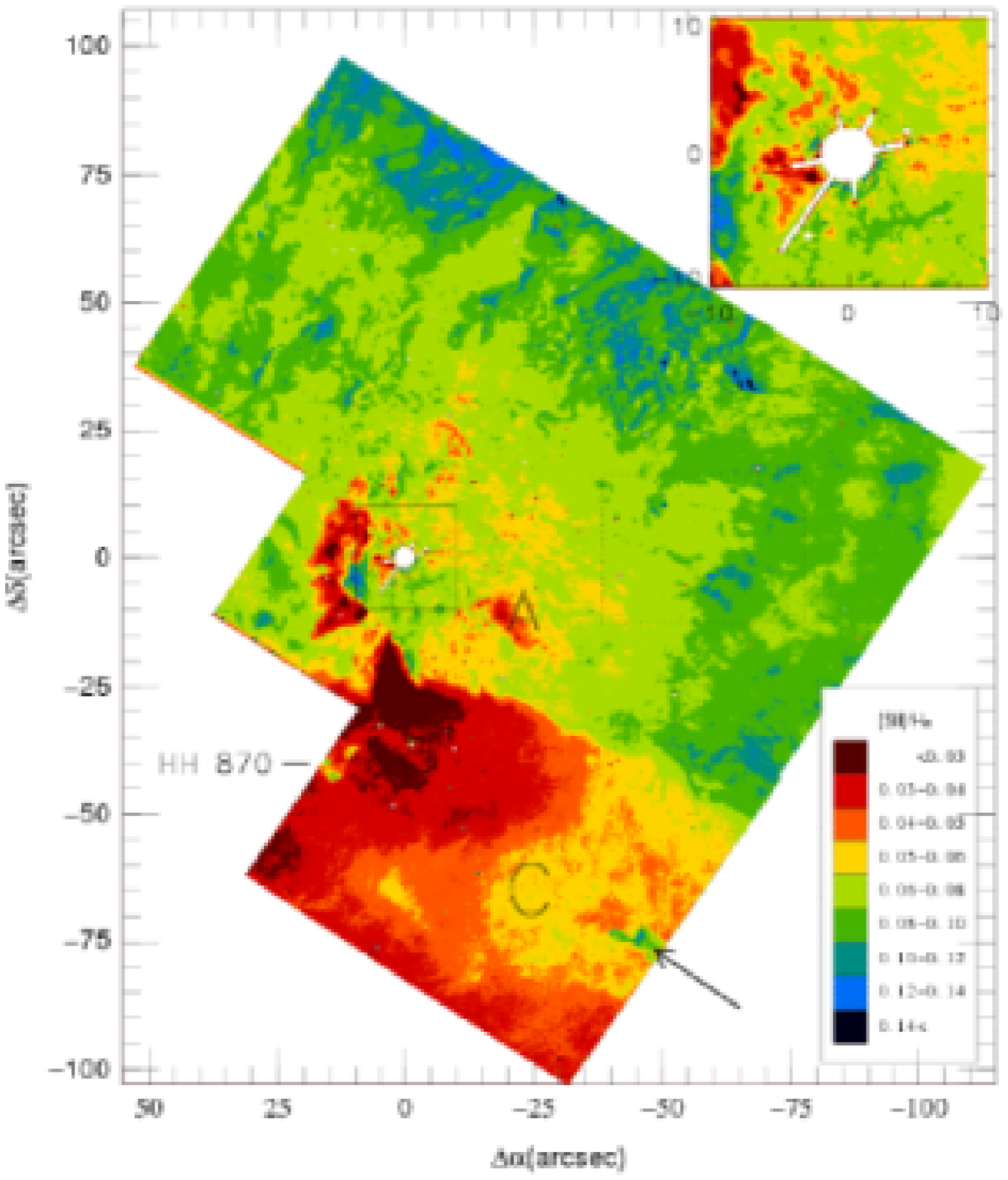}
\caption{[S\,{\sc ii}]/H$\alpha$ (S2H) ratio map of the Hourglass region.
An enlargement of the central part is shown in the upper right corner of 
the panel. The letters A, B and C refer to the regions described in the
text (see sections~\ref{dist} and \ref{signatures}). The dashed-lined 
rectangular box defines the area shown in Fig.~\ref{hh_ring}. 
The arrow indicates the direction to the O-type
star 9\,Sgr, located 4\farcm12 from the finger-shaped molecular globule.
}
\label{s2ha_map}
\end{figure*}

Figures~\ref{s2_map} and \ref{s2ha_map} show the {\em HST} [S\,{\sc ii}]
continuum-subtracted emission map and the corresponding
[S\,{\sc ii}]/H$\alpha$ (S2H) ratio map of the Hourglass Nebula,
respectively.
The strongest line emission arises from the inner part of the
Hourglass and is mainly concentrated on the east wall of the cavity.
Both, the [S\,{\sc ii}]
and the H$\alpha$ emissions, have a clumpy structure close to Her\,36. 
When compared to H$\alpha$ emission, the [S\,{\sc ii}] emission shows a more 
complex and filamentary pattern to the north and west of the nebula.
On the other hand, both images look smoothed in the southern part of the 
nebula, where the H\,{\sc ii} cavity is an opened blister.
These differences in the emission appears as 
large fluctuations in the S2H map, in a scale of few tenths of arcsecond,
especially close to Her\,36 and to the west part of the image.

It is very interesting to note that the S2H flux ratio goes below 0.10 
over almost the whole region of the Hourglass, 
as expected for a photoionized plasma. 
We are particularly interested in detecting emission features produced
in shock-excited gas, for which the expected S2H flux ratio values are
typically of 0.3--0.5. Nevertheless, the highest S2H flux ratio 
hardly reaches 0.2 (over an arc located at 75'' to the NW of Her\,36). 
This could be explained by considering {\em normal} photoionized gas filling 
the volume in front of the nebula. 
This gas shows a S2H flux ratio close to 0.05, and then 
reduces the contrast between the shock-excited features with
higher values and the molecular cloud. 
As mentioned in Section~\ref{dist}, 
the dust associated with this foreground material 
could be responsible for most of the foreground extinction in the nebula.

\begin{figure*}
\includegraphics[width=180mm]{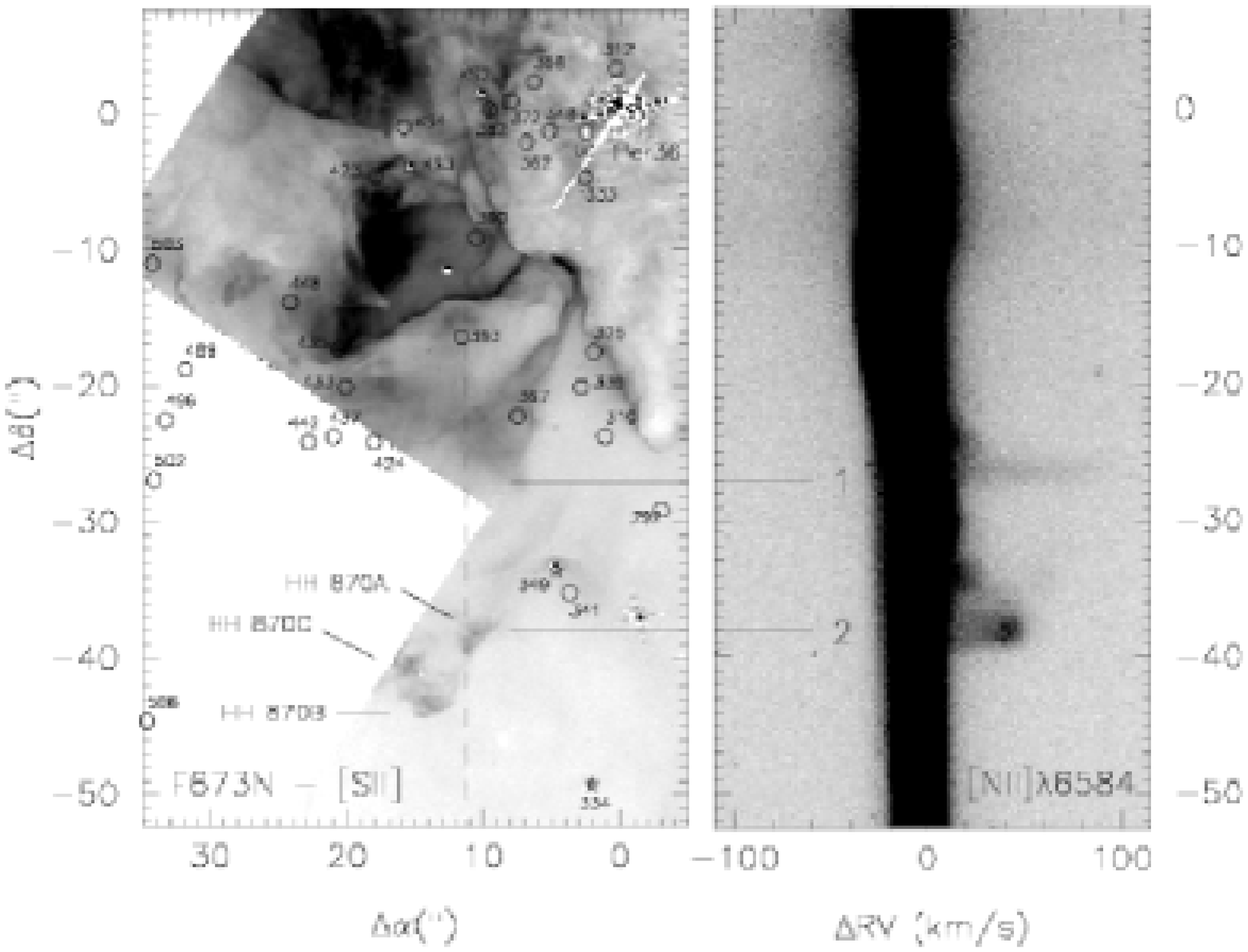}
\caption{Right-hand panel: long-slit spectrum in the [N\,{\sc ii}] 6584\,\AA\ 
emission line obtained across the Hourglass Nebula with NTT. 
Left-hand panel: [S\,{\sc ii}] WFPC2 mosaic showing the location of the 
long-slit aperture (dashed line) and the infrared sources (circles).
The stars indicate the position of the two sources showing bow shocks 
around them.}
\label{hh_slit}
\end{figure*}

Perhaps the most stricking features in the [S\,{\sc ii}] and S2H images
are three nebular knots located 40'' to the SE of Her\,36 
that appear to form one large structured bow shock for which the HH~number
HH\,870 has been assigned in the HH~catalogue\footnote{HH~catalogue numbers are
assigned by B.~Reipurth in order to correspond with the list of Herbig-Haro
objects that he maintains.}.
Equatorial positions for this and the other new HH~objects discovered 
in the Hourglass region are shown in Table~\ref{HH}.  
The morphological appearance of HH\,870 is
pretty similar to the HH\,203 and HH\,204 objects in the Orion Nebula 
(O'Dell et al. 1997).
Its three flow components have been labelled as {\it A}, {\it B} and {\it C}
in Figure~\ref{hh_slit}.  
The right-hand panel of 
this figure shows a long-slit spectrum in the 
[N\,{\sc ii}] 6584\,\AA\ emission line obtained across the Hourglass Nebula
with NTT. The placement of the long-slit aperture has been plotted on the 
[S\,{\sc ii}] WFPC2 mosaic in the left-hand panel of the same figure.
Two noticeable kinematic structures (labeled as ''1'' and ''2'')
are clearly observed. These velocity components are associated with nebular 
structures located at 25'' (0.15 pc) and 37'' (0.22 pc) from the middle point 
of the slit. 
Their receding velocities respect to the H\,{\sc ii} region reach 
80~km\,s$^{-1}$ for feature ``1'' and 
45~km\,s$^{-1}$ for feature ``2''.
Feature ''1'' seems to 
origin in the gas located inmediately to the south of a dusty finger-like 
structure whose head points directly to Her\,36.
Feature ''2'' is clearly associated with the flow component
{\it A} of HH\,870. 
It must be taken in mind that the long-slit images were originally obtained 
with other scientific purposes, and so the position on the sky was chosen.
It was a fortunate coincidence that the slit aperture crossed the western 
bow shock feature, but
it is presumably not aligned along the jet axis. This could explain the
undetection of the northern counterjet, although
it also might be due to the fact that the driven source is buried in the 
molecular cloud core. 
Both infrared sources \#414 (KS\,4) and \#410 (KS\,3) are located along the
simmetry axis of the main bow shock component {\it B} of HH\,870. 
As was remarked in
Section~\ref{individual}, KS\,4 is probably a Class~I protostar and therefore
a good candidate to be the driving source of this structure. 

\begin{table*}
\centering
\caption{Equatorial coordinates of the new HH~objects in the Hourglass Nebula}
\label{HH}
\begin{tabular}{cccl}
\hline\\
Object   &  $\alpha$(J2000.0)  &  $\delta$(J2000.0)    &   Comments \\
         &                     &                       &            \\
\hline\\
HH\,867   &  18:03:36.77        &  -24:22:33.0          & $4''\times2\farcs5$ ring, strong in [S\,{\sc ii}] \\     
HH\,868   &  18:03:36.00        &  -24:22:49.0          & Three emission arcs, 6'' in size            \\  
HH\,869   &  18:03:35.69        &  -24:22:30.3          & Strong [S\,{\sc ii}] knot           \\    
HH\,870   &  18:03:41.44        &  -24:23:25.0          & Three strong emission knots, 10'' in size            \\
         &   &    &   \\             
\hline\\                        
\end{tabular}
\end{table*}

\begin{figure}
\includegraphics[width=95mm, angle=270]{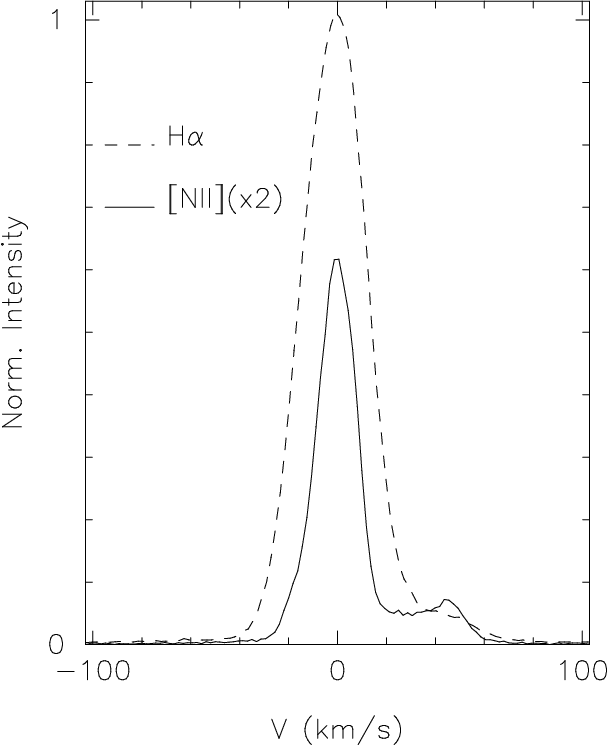}
\caption{Superposition of the H$\alpha$ and [N\,{\sc ii}] 6584\,\AA\ 
(scaled by 2) profiles along the position of feature "2" in 
Figure~\ref{hh_slit}.}
\label{hh_spec}
\end{figure}

One-sided outflows are commonly observed emerging from dense cores 
in the photoionized medium (cf. HH\,616 and HH\,617 in S140,
Bally et al. 2002; HH\,777 in IC\,1396N, Reipurth et al. 2003).
Additional evidence 
in favour that kinematic feature ``2'' (spatially associated with the 
feature HH\,870 {\it A}) is a bona fide bow shock is found in the very high 
[N\,{\sc ii}]/H$\alpha$ (N2H) ratio  derived from the long-slit spectra. 
Figure~\ref{hh_spec} shows the superposition of the
H$\alpha$ and [N\,{\sc ii}] 6584\,\AA\ (scaled by 2) profiles 
along the position of feature "2". 
The N2H ratio value is about 0.15 for the main emission component,
but increases to about 0.7 for the radial velocity component at 
+45~km\,s$^{-1}$ , 
reinforcing the idea that this feature is in fact shock-excited.

\begin{figure*}
\includegraphics[width=105mm]{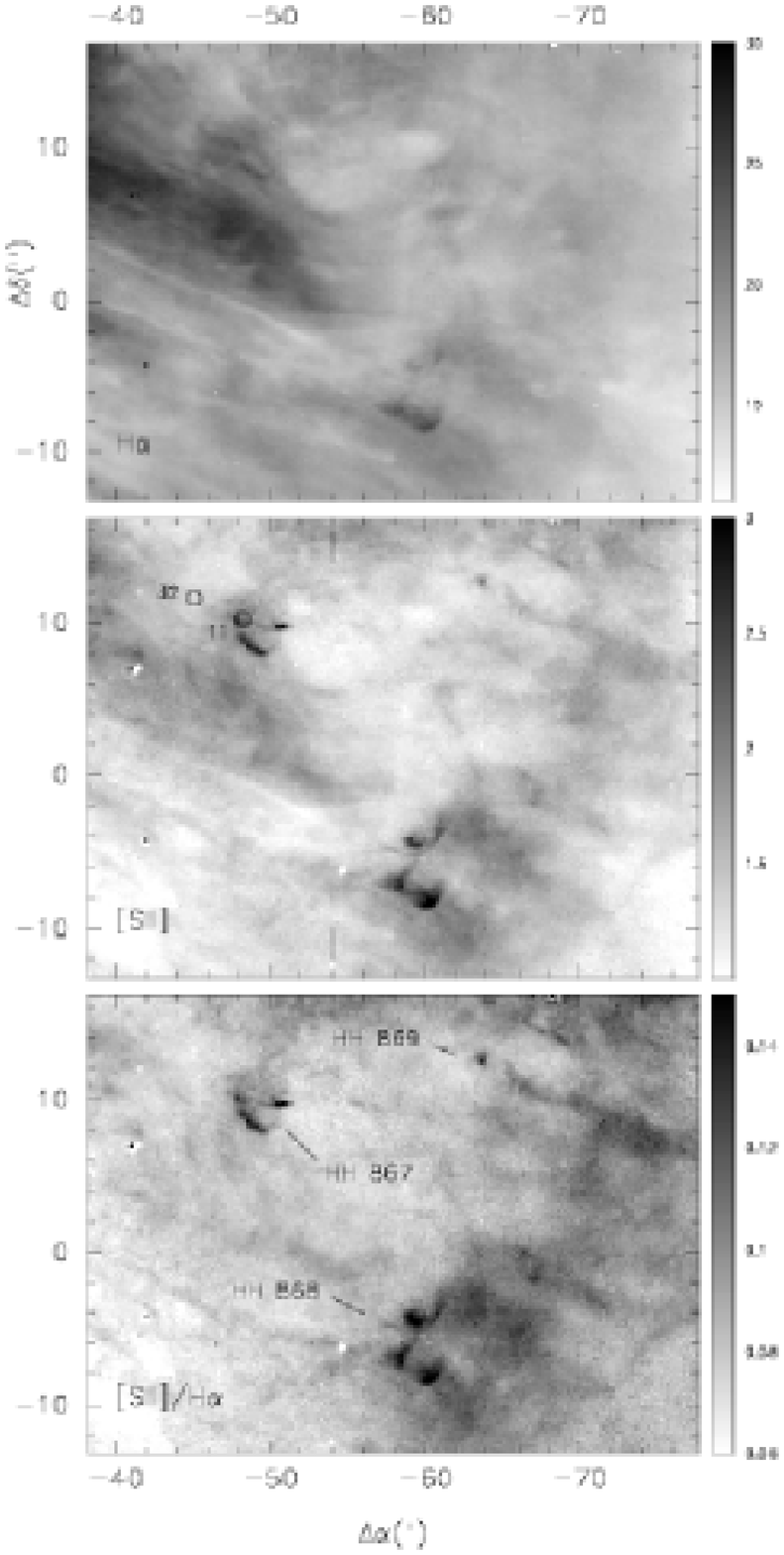}
\caption{ WFPC2 H$\alpha$ and [S\,{\sc ii}] mosaics (top and middle panels)
and S2H ratio map (bottom panel)
showing three of the new HH~objects identified in the Hourglass region
(HH\,867, HH\,868 and HH\,869) 
The infrared sources \#32 and \#11 are also marked with circles. For the top
and middle panels, the surface brightness units are 
$10^{-16}$\,erg cm$^{-2}$ s$^{-1}$ per WFC pixel.}
\label{hh_ring}
\end{figure*}

About 50'' (0.3 pc) to the West of Her\,36, we find another set of nebular 
structures which are also identified as new HH~objects.
Figure~\ref{hh_ring} shows a portion of the WFPC2  
H$\alpha$ and [S\,{\sc ii}] mosaics and the S2H ratio map, 
where these interesting features have been labelled as HH\,867, HH\,868 and 
HH\,869 according to the numbers assigned in the HH~catalogue.

The object HH\,867 is a very intriguing elliptical ring, approximately
$4''\times2\farcs5$ in size ($5000\times3100$\,AU).
The ring is uncomplete and presents strong emission in [S\,{\sc ii}]
and much weaker emission in H$\alpha$, leading to a S2H ratio of about 0.11.
A jet-like feature seems to cross axially the center of the ring
with a certain inclination angle respect to the ring plane.
This candidate jet shows two main condensations with high S2H ratio:
the first one (S2H$\sim$0.18) is placed toward the west of the ring, 
whereas the other (S2H$\sim$0.11) locates on its eastern border.
Furthermore, the jet seems to extend beyond the west end of the ring 
in a diffuse path of about 28'' (0.17 pc),
which can be mainly seen in the S2H ratio map.
Near the apparent jet axis, there are two infrared excess sources, \#32 
($K_s=13.04$, $J-H=1.50$, $H-K=1.00$) and \#11 
($K_s=15.73$, $J-H=1.14$,$H-K=1.25$), which seem to be connected by a thin 
H$\alpha$ filament. Source \#449 coincides with the eastern jet condensation.
Since both of these sources are candidates to be embedded YSOs,
any of them might be the driving source of the jet feature.

HH\,868, located about 20'' to the SE of HH\,867, is an ever larger
structure (6'' in size = 7500 AU)
composed by three nebular emission arcs. 
These bow shock features show  high S2H ratio values ($\sim0.13$) and they are 
undetectable in the H$\alpha$/H$\beta$ map, which means that the extinction 
has foreground origin. The northern arc shows a bright [S\,{\sc ii}]
condensation inside the bow show, suggesting a reverse shock
(''Mach disk''), indicative of the action of a faster jet overtaking cooled 
gas behind the preceding bow shock.
Note that it seems very likely based on the morphology of the flow
components that HH\,867 and HH\,868 form independent bow shocks in a single 
large flow coming from the upper left of the image.

A third HH~object is found in Fig.~\ref{hh_ring}.
HH\,869 is a bright [S\,{\sc ii}] condensation located 15'' (19000 AU)
to the west of the HH\,867, almost perpendicularly to the previously
mentioned ring plane. 
This condensation is not appreciable in H$\alpha$ emission, 
showing a S2H ratio value 
of $\sim0.12$. Our infrared survey does not reach this peculiar emission 
feature and the 2MASS catalog does not show any infrared source close to this 
location. A diffuse filament seems to 
develop from HH\,869, grow in the southwest direction and intersect the 
filament coming from HH\,867. 
Bally et al. (2002) describe the breakout in an ionized medium of a jet 
arised inside the molecular cloud. They found that the wall of the molecular 
cloud shows arcs with enhanced [S\,{\sc ii}] emission where the jet appears.
In this way, we should also take in mind the 
hypothesis that the ring of HH\,867 represents the point in the molecular 
cloud 
where a jet arised in (or around) Her\,36 emerges, since the normal direction 
to the ring points directly to that star. Furthermore, the HH\,869 feature 
is placed along the line joining the ring and Her\,36.

Chakraborty \& Anandarao (1997, 1999) performed Fabry-Perot observations of 
the Hourglass region in the [N\,{\sc ii}] 6583\AA\  and [O\,{\sc iii}] 5007\AA\ emission 
lines. They found high expansion velocities up to 50 km\,s$^{-1}$, which they 
interpreted as indicative of Champagne flows. 
Unfortunately, the area observed by these authors does not reach the 
location of the outflow HH 870.
Nevertheless a careful inspection of their plots suggests 
the possibility that some of the high velocity features observed in their
spectral line profiles may represent the contribution of collimated 
outflows to the general turbulent motion.
For example, among the line profiles shown in Figure 2 of 
Chakraborty \& Anandarao (1997),
there is a high velocity component of -50 km\,s$^{-1}$, which appears very 
prominent at P.A. 190$^\circ$, 20'' from Her 36 (Fig. 2h) 
but it is not observed
at P.A. 175$^\circ$, 12'' from this star (Fig. 2f).
This kind of small-scale variations in the structure of the velocity field 
may be indicating the existence of collimated outflows.
  
Located 30'' (0.18 pc) to the NW of Her\,36, source \#156 (SCB\,1040) 
is a remarkable object which has already been suggested as 
pre-main sequence candidate 
from its optical colours in {\em HST} data by Sung et al. (2000).
Its H$\alpha$ image shows extended emission ($0\farcs5$ in size)
and reveals a chain of nebular emission features extending 10'' to the south, 
which clearly resembles the 
one-sided jets detected in the Orion Nebula (Bally et al. 2000).

About $7\farcs5$ (9400~AU) to the South of source \#156 is source \#144, 
which shows an intense and very elongated H$\alpha$ emission. This star has 
infrared colours similar to T~Tauri stars ($J-H=1.36$, $H-K_s=0.82$). 
Additionally, its $K_s$ magnitude of 10.14 corresponds to
the same brightness as T~Tauri itself ($J-H=1.02$, $H-K_s=0.91$ 
and $K_s=5.33$, according to 2MASS Second Incremental Data Release, 
Cutri et al. 2000), if moved from its location in the Taurus~T association 
(140~pc) to the distance of M8 (1300~pc).

A rich clustering of faint infrared sources (\#369, 374, 497, 320, 370)
is found 35'' (0.2~pc) to the North of Her\,36, close to bright
filaments facing 9\,Sgr. These filaments are part of the PDR of the molecular
core detected in CO (see Figure~\ref{colores}).

The central part of the Hourglass Nebula shows an incredible complex 
structure which will be better described in a subsequent paper.
We remark here a large dust pillar orientated to Her\,36 in the SE-NW 
direction, which is observed as an extinction feature against the nebular 
background. Its very bright rimmed head indicates that
it is presumably illuminated on its far side, and so is somewhat in front 
of Her\,36. The projected distance
between the head of the pillar and the star is 11'' (0.07~pc).
Our $K_s$-band image reveals a bright stellar infrared source (\#347) 
located in the tip of this finger, resembling those found in 
the dense knots of 30~Doradus by Walborn et al. (1999).
Source \#347 has no optical counterpart and its near-infrared colours are 
typical of a highly reddened early-type star ($J-H=1.72$, $H-K_s=1.05$).

In the southern region of the field, the WFPC2 images reveal another 
interesting dust structure, which might probably be an externally-ionized 
molecular globule similar to that found in the core of the Carina Nebula by 
Smith et al. (2004).
This molecular globule, located $\sim105''$ to the SW of the Hourglass,
has also a shape resembling a human finger that points
toward its likely source of ionizing photons. Curiously, the finger does not
point to Her\,36 but to the O4\,V((f))-type star 9\,Sgr.
While the finger is seen only as a silhouette against a brighter background in 
H$\alpha$, its surface is very bright in [S\,{\sc ii}], 
which is especially clear from the S2H ratio map in Figure~\ref{s2ha_map}.
Although most of this structure is out of the FOV of our infrared images, 
a very bright infrared stellar source is noticeable in the summit of the 
pillar. 2MASS archive data reveal the presence of a highly reddened object at
this location ($J=14.603$, $J-H=3.018$, $H-K_s=1.502$).

Finally we have identified two large bow shocks surrounding young stars
whose near-infrared colours are typical of reddened T~Tauri stars 
(sources \#349, $J-H=1.25$, $H-K_s=0.64$, and \#334, $J-H=1.44$,
$H-K_s=0.84$; see Fig.~\ref{hh_slit}). 
These objects are marked with stars in Figure~\ref{hh_slit}.
It is interesting to note that, while source \#334 is a single star in 
{\em HST} images, it is observed as a double star with components separated 
by 0\farcs5 in our infrared images. 
The eastern component is not detected in the optical images, but it
is brighter than its companion in the infrared.
This object seems to be a resolved binary with components affected
by a different degree of extinction.

The bow shocks detected in sources \#349 and \#334
are seen as arcs consisting in bright H$\alpha$ emission and facing Her\,36. 
In both cases, they are concave toward the corresponding
infrared source and convex toward Her\,36.
The bow shock around star \#349 is brighter than the one around star \#334.
Their morphologies resemble strongly the hyperbolic bow shock around
the T~Tauri star LL~Ori (Bally et al. 2000, Bally \& Reipurth 2001).
Thus, these bow shocks are candidates to be wind-wind collition fronts.
 
In summary, the morphological analysis of the {\em HST} images of the 
Hourglass region reveals that the nebula has a complex structure, 
showing dozens of features that closely resemble those observed in other star 
forming regions (for example, M42, M16, NGC~3372). However, 
almost nothing is known yet about the proper motion and radial velocity
of these objects, so it is premature to draw any conclusions about these
objects here.   
In a forthcoming paper, we will discuss optical spectroscopy recently
obtained using the Boller and Chivens Spectrograph 
on the 6.5\,m Magellan I telescope at Las Campanas Observatory,
which will undoubtedly contribute to unveil their true nature.

\section{Summary and Conclusions}

We have obtained near-infrared $JHK_s$ images of the Hourglass Nebula in 
M8. These data were complemented with {\em HST} images and longslit 
spectroscopy retrieved from on-line databases. 
The main results can be summarized as follows:

We used the recently developed numerical code CHORIZOS to obtain a new
estimate for the distance to the Hourglass region. Using optical and infrared 
photometry from this work and from literature for a group of early-type stars 
around Herschel\,36 (including this star), 
we derived a distance modulus of 10.5 (1.25 kpc). 
This value is sensibly smaller than the 11.25 previously accepted for the 
region, but it is in excellent agreement with the last 
determination for NGC~6530 by Prisinzano et al. (2005).

Our infrared images confirm the existence of a very young stellar cluster
around the massive O-type star Herschel\,36. We have detected almost 100
sources with colours indicative of intrinsic infrared excess emission, 
which appear as prime candidates to be T~Tauri and Herbig Ae/Be stars. 
These objects are not uniformly distributed in the region, but they 
extend along the molecular core, mostly concetrated near the two most
intense peaks of the CO distribution (White et al. 1997).
The large fraction of infrared excess sources (over 70~\% in the central part
of the field) along with the evidence of outflow activity from Her\,36
(Burton 2002), suggest the stellar population within the cluster
must be extremely young ($\sim 10^6$\,years).

By comparing the $JHK_s$ photometric diagrams of the Hourglass and a 
2MASS control field, we have found that a significant fraction of the
stars observed toward M8 may be red giants belonging to the galactic
inner disk and Bulge.  
Based on the method described by Lada et al. (1994),
we have used the positions and near-infrared colour excess estimates of 
over 400 background field giants to directly measure the extinction through 
the molecular cloud, finding a strong correlation 
with the distribution of molecular material.

Archival {\em HST} emission-line images reveal a variety of ongoing star 
formation features in the core of M8.
Several of the candidate YSOs detected here
seem to be associated to structures similar to those observed in other
star forming regions like M16 and M42, namely,
proplyds, jets, dense knots, molecular globules and bow shocks. 
Furthermore, four Herbig-Haro objects have been detected for the first time
in the region (HH\,867, HH\,868, HH\,869 and HH\,870).
A longslit spectrum in the [N\,{\sc ii}] 6584\,\AA\ emission line obtained 
across the Hourglass Nebula with NTT and retrieved from the ESO Archive 
Facility confirmed the identification of HH\,870, 
providing the first direct evidence of active star formation
by accretion in M8.
In order to understand better the nature of some of the rest of the objects
we have obtained optical spectroscopy with the 6.5\,m Magellan Telescope,
which will be analyzed in a future paper.

\section*{acknowledgements}

This publication makes use of data products from the Two Micron All Sky
Survey, which is a joint project of the University of Massachusetts and the 
Infrared Processing and Analysis Center/California Institute of Technology, 
funded by the National Aeronautics and Space Administration and the National 
Science Foundation. 
This research has made use of Aladin and the Simbad Database, operated at CDS,
Strasbourg, France.

Financial support from ``Fondo de Publicaciones de la
Direcci\'on de Investigaci\'on de la Universidad de La Serena (Chile)''
and from the Center for Astrophysics (FONDAP No. 1501003, Chile) are
acknowledged by RHB and MR, respectively. 
JIA thanks the Departamento de F\'{\i}sica of Universidad de La Serena for the
use of their facilities and the warmest hospitality.
Also, the authors gratefully thank the staff at LCO for kind hospitality 
during the observing run.

\end{document}